\documentclass[fleqn,usenatbib]{mnras}

\usepackage{newtxtext,newtxmath}
\usepackage{threeparttable}
\usepackage{xspace}%

\usepackage[T1]{fontenc}
\usepackage{ae,aecompl}

\usepackage{graphicx}	
\usepackage{amsmath}	
\usepackage{amssymb}	

\newcommand{\codename}{\texttt{juliet}}
\newcommand{\planetnameb}{TOI-141b}
\newcommand{\planetradiusb}{$1.745^{+0.051}_{-0.052}$}
\newcommand{\planetmassb}{$8.83^{+0.66}_{-0.65}$}
\newcommand{\planetrhob}{$9.15^{+1.1}_{-1.0}$}
\newcommand{\planetgravb}{$28.5^{+2.8}_{-2.7}$}
\newcommand{\planetTeqb}{$2128^{+13}_{-14}$}
\newcommand{\planetab}{$0.02012^{+0.00015}_{-0.00012}$}
\newcommand{\planetnamec}{TOI-141c}
\newcommand{\planetmassc}{$19.95^{+1.38}_{-1.36}$}
\newcommand{\planetTeqc}{$1265.4^{+7.3}_{8.4}$}
\newcommand{\planetac}{$0.056798^{+0.00044}_{-0.00032}$}
\newcommand{\starname}{TOI-141}

\newcommand{\rsolid}{$r_{\rm core+mantle}$\xspace}

\newcommand{\fesima}{{\rm Fe}/{\rm Si}_{\rm mantle}\xspace}
\newcommand{\mgsima}{{\rm Mg}/{\rm Si}_{\rm mantle}\xspace}

\title[HD 213885b (\planetnameb) and HD 213885c (\planetnamec)]{HD 213885b: A transiting 1-day-period super-Earth with an Earth-like composition around a bright ($V=7.9$) star unveiled by TESS}

\author[Espinoza et al.]{N\'estor Espinoza$^{1,2}$\thanks{E-mail: espinoza@mpia.de (NE)}\thanks{Bernoulli Fellow}\thanks{IAU-Gruber Fellow},
Rafael Brahm$^{3,4,6}$,
Thomas Henning$^{2}$,
Andr\'es Jord\'an$^{5,6}$,
\newauthor
Caroline Dorn$^7$,
Felipe Rojas$^{4,5}$,
Paula Sarkis$^{2}$,
Diana Kossakowski$^{2}$, 
Martin Schlecker$^{2}$,
\newauthor
Mat\'ias R. D\'iaz$^8$,
James S. Jenkins$^8$, 
Claudia Aguilera-Gomez$^9$,
Jon M. Jenkins$^{10}$,
\newauthor
Joseph D. Twicken$^{11}$, 
Karen A. Collins$^{12}$, 
Jack Lissauer$^{10}$, 
David J. Armstrong$^{13,14}$,
\newauthor
Vardan Adibekyan$^{15}$,
David Barrado$^{16}$,
Susana C. C. Barros$^{15}$,
Matthew Battley$^{13,14}$,
\newauthor
Daniel Bayliss$^{13,14}$,
Fran\c{c}ois Bouchy$^{17}$,
Edward M. Bryant$^{13,14}$,
Benjamin F. Cooke$^{13,14}$,
\newauthor
Olivier D. S. Demangeon$^{15}$,
Xavier Dumusque$^{17}$,
Pedro Figueira$^{18,14}$,
Helen Giles$^{17}$,
\newauthor
Jorge Lillo-Box$^{18}$,
Christophe Lovis$^{17}$,
Louise D. Nielsen$^{17}$,
Francesco Pepe$^{17}$,
\newauthor
Don Pollacco$^{13}$,
Nuno C. Santos$^{15,19}$,
Sergio G. Sousa$^{15}$,
St\'ephane Udry$^{17}$,
\newauthor
Peter J.\ Wheatley$^{13,14}$,
Oliver Turner$^{17}$,
Maxime Marmier$^{17}$,
Damien S\'egransan$^{17}$,
\newauthor
George Ricker$^{20}$, David Latham$^{12}$, 
Sara Seager$^{20}$,
Joshua N. Winn$^{21}$,
John F.\ Kielkopf$^{22}$, 
\newauthor
Rhodes Hart$^{23}$, 
Geof Wingham$^{24}$, 
Eric L.\ N.\ Jensen$^{25}$,
Krzysztof G.\ He{\l}miniak$^{26}$,
\newauthor
A. Tokovinin$^{27}$,
C. Brice\~no$^{27}$, 
Carl Ziegler$^{28}$,
Nicholas M. Law$^{29}$, 
Andrew W. Mann$^{29}$,
\newauthor
Tansu Daylan$^{19}$\thanks{Kavli Fellow},
John P. Doty$^{30}$, 
Natalia Guerrero$^{19}$, 
Patricia Boyd$^{31}$, 
Ian Crossfield$^{20}$,
\newauthor
Robert L. Morris$^{11,10}$,
Christopher E. Henze$^{10}$,
Aaron Dean Chacon$^{32,10}$
\\
The authors' affiliations are shown in Appendix \ref{aa}.
}

\date{Accepted XXX. Received YYY; in original form ZZZ}

\pubyear{2015}

\begin{document}
\label{firstpage}
\pagerange{\pageref{firstpage}--\pageref{lastpage}}
\maketitle

\begin{abstract}
We report the discovery of the 1.008-day, ultra-short period (USP) 
super-Earth HD 213885b (\planetnameb) orbiting the bright ($V=7.9$) star HD 
213885 (TOI-141, TIC 403224672), detected using photometry from the recently launched \textit{TESS} mission. Using FEROS, 
HARPS and CORALIE radial velocities, we measure a precise mass of 
$8.8 \pm 0.6 M_\oplus$ for this $1.74 \pm 0.05 R_\oplus$ 
exoplanet, which provides enough information to constrain 
its bulk composition --- similar to Earth's but enriched in iron. The radius, 
mass and stellar irradiation of HD 213885b are{, given our data, very similar to} 55 Cancri e, making this exoplanet 
{a good target to perform comparative exoplanetology of short period, highly irradiated super-Earths.} 
Our precise radial velocities 
reveal an additional $4.78$-day signal which we interpret as 
arising from a second, non-transiting planet in the system, HD 213885c, whose minimum mass 
of $19.9 \pm 1.4 M_\oplus$ makest it consistent with being a Neptune-mass exoplanet. The HD 213885 system is 
very interesting from the perspective of future atmospheric 
characterization, being the second brightest star to host an 
ultra-short period transiting super-Earth (with the brightest star being, in fact, 
55 Cancri). Prospects for 
characterization with present and future observatories are discussed.
\end{abstract}




\section{Introduction}
The successfully launched and currently operating Transiting 
Exoplanet Survey Satellite \citep[TESS,][]{tess} is set to 
become one of the most important missions in the search for 
small, characterizable rocky exoplanets. Currently 
exploring almost the whole sky on the hunt for 
transiting exoplanets orbiting bright ($V<13$) stellar 
hosts, TESS' primary mission is to generate a sample of 
small ($<4R_\oplus$) exoplanets for which precise masses 
and even atmospheric characterization will be 
possible, revolutionizing our view of these small, 
distant worlds.

Among the distinct populations of small exoplanets, one of 
the most interesting are the so-called 
Ultra-Short-Period (USP) exoplanets. These are 
planets that orbit at extremely short periods ($P\leq 1$ day), 
smaller than about $2R_\oplus$, and which appear to 
have compositions similar to that of the Earth \citep{USP}. 
Although {almost} a hundred of these systems have been 
found by the \textit{Kepler} mission, with which it was found that 
these exoplanets are extremely rare \citep[about as 
rare as hot-jupiters,][]{USPOC}, only a handful of them 
have precise radii and masses, as the stars in the 
\textit{Kepler} field are typically much too faint for 
spectroscopic follow-up. Transit surveys like TESS, however, 
are the perfect haystacks to find these rare needles as 
they are designed to find short-period transiting exoplanets 
around bright stellar hosts, allowing us to explore the yet poorly 
understood dimension of mass and, thus, bulk composition 
of these interesting extrasolar worlds. In addition, missions 
like TESS are extremely important for exoplanets such as 
USPs as they will generate a sample of them which will be 
prime targets for future atmospheric follow-up with missions 
like the upcoming James Webb Space Telescope (JWST), which 
will in turn allow us to explore the exciting dimension 
of atmospheric composition of these small, short-period exoplanets \citep[see, e.g., the case of 55 Cancri e;][]{Demory:2016,Angelo:2017,Miguel:2019}.

The possibility to perform spectroscopic follow-up for these 
USPs is in turn also interesting because of another 
fact: the inclination between the orbits of multi-planetary 
systems appears to be larger for short-period exoplanets in 
tight orbits \citep[$\Delta i = 6.7 \pm 0.6$ degrees for 
planets with $a/R_* \leq 5$ versus $2.0 \pm 0.1$ for planets with $5<a/R_*<12$,][]{MI:2018}, which might be a signature of 
orbital migration due to excitation effects such as high-eccentricity migration \citep{Petrovich:2018}. If this 
effect is indeed the one dominating in systems having USPs, {then detecting transits of more than one planet} 
in multi-planetary systems might be intrinsically harder {to do} than for systems not having them, 
as the increased mutual inclination between the exoplanets in 
the system might prevent us from observing the transits of the 
other members of it. However, if their inclinations are within 
the same order of magnitude, these extra members might be 
found via high-precision spectroscopic follow-up, and this 
might in turn provide valuable constraints on the 
mutual inclinations between the exoplanets of these systems that 
might aid in the understanding of the formation of these 
rare, small exoplanets. 

In this work, we present the discovery and characterization 
of \textbf{a new} USP discovered by the TESS mission, HD 213885b (\planetnameb), 
characterized thanks to precise radial-velocity measurements 
from FEROS, HARPS and CORALIE. In addition to the tight 
constraint on the mass of this new exoplanet, our 
radial-velocity measurements reveal the presence of an additional non-transiting exoplanet in 
the system, HD 213885c (\planetnamec).

We organize this work as follows. In Section 
\ref{sec:data} we present the data used to make the discovery 
of this multi-planet system. In Section \ref{sec:analysis} we 
present the analysis of this data, in which we derive the 
properties of both the star and the planets in the system. 
In Section \ref{sec:discussion} we present a discussion on the 
system and the implication of this discovery to both 
the overall population of small exoplanets and the known 
USPs and in Section \ref{sec:conclusions} we present the 
conclusions of our work.

\section{Data}
\label{sec:data}
\subsection{TESS photometry}
\begin{figure*}
   \includegraphics[height=0.7\columnwidth]{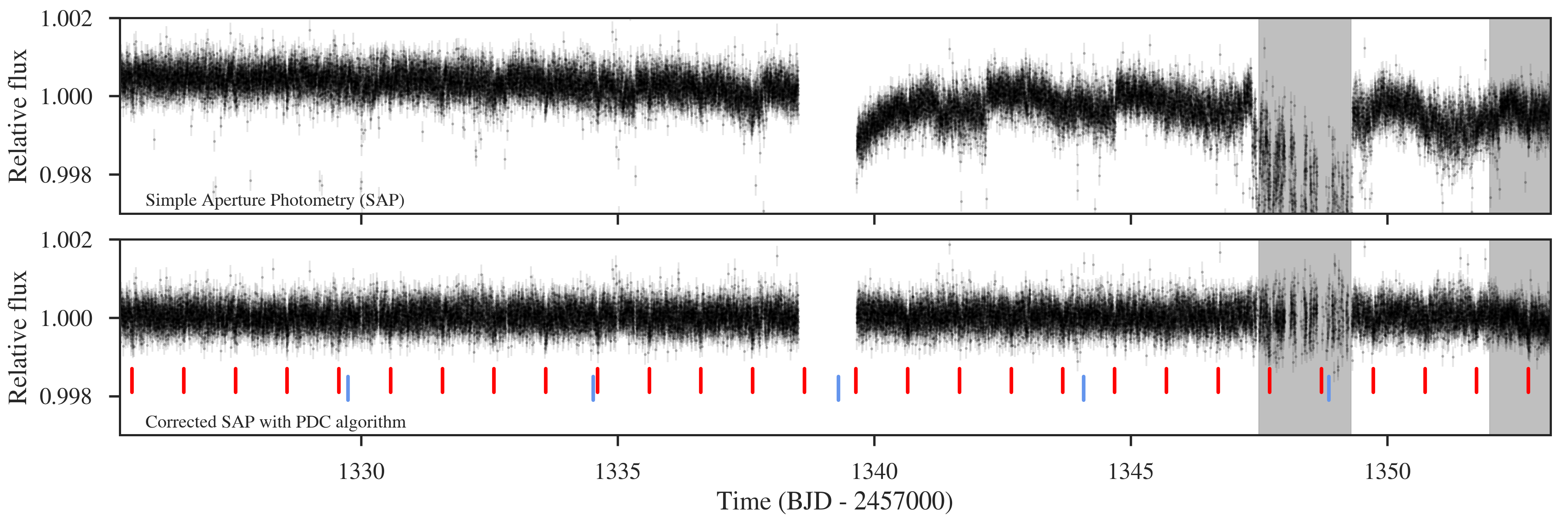}
   \includegraphics[height=0.7\columnwidth]{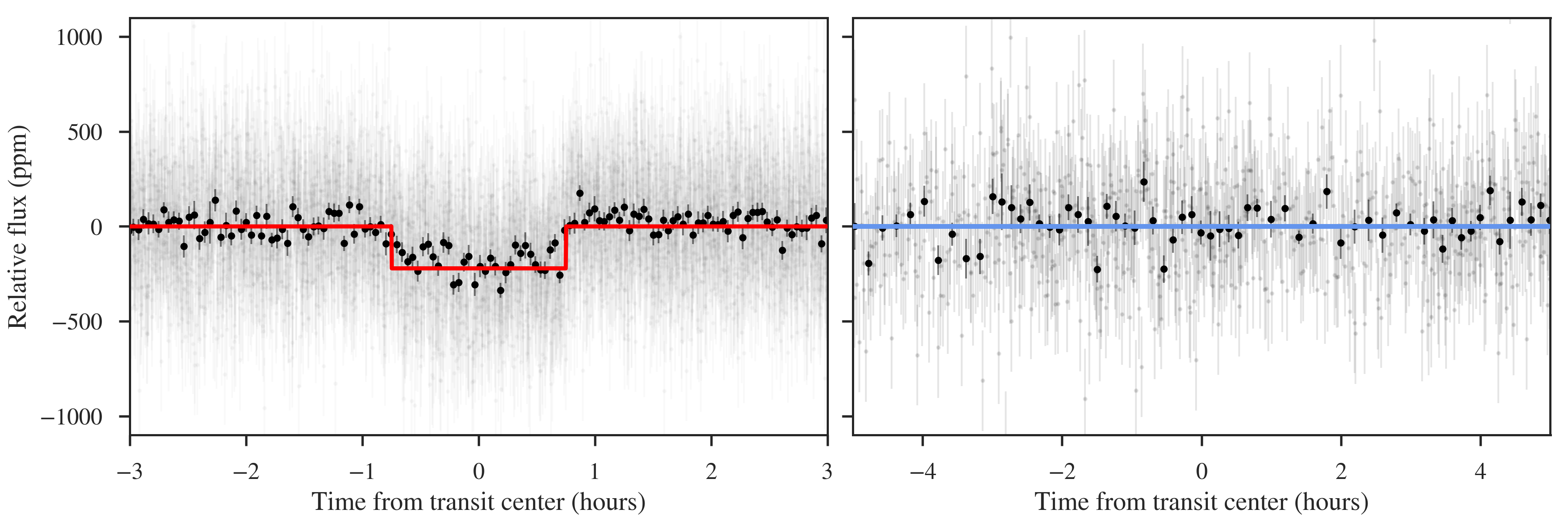}
    \caption{{Top panels.} TESS photometry for \starname. The upper panel shows the 
    Simple Aperture Photometry (SAP) provided by the TESS alerts 
    (\texttt{SAP\_FLUX}) along with the corresponding errors (\texttt{SAP\_FLUX\_ERR}) after normalizing for the median flux. 
    The same photometry but systematics-corrected via the PDC algorithm, 
    also provided by the TESS alerts (\texttt{PDC\_SAP\_FLUX}) is shown 
    in the bottom panel along with the corresponding errors (\texttt{PDC\_SAP\_FLUX\_ERR}). Red lines indicate transits of the 1-day planetary candidate \starname.01 for which 
    (200 ppm) transits can be easily observed by eye in the photometry. 
    Blue lines indicate the expected position of transits of a 
    second, $4.75$-day sinusoidal signal found on the radial-velocity measurements 
    (see Section \ref{sec:jointanalysis}) --- no transits are evident at 
    those times. Grey regions indicate portions of the time-series left out of our analysis (see text). {Bottom panels.} Phased photometry at interesting periods. The left-most phased 
    photometry shows the photometry phased at the period of TOI-141.01 (grey points); black points show binned datapoints for visualization. Red line indicates the box model implied by our BLS search. The {right} panel shows the same for the $4.75$-day sinusoidal signal found in our radial-velocity measurements (see Section \ref{sec:jointanalysis}), where the 
    reported time of transit-center is the expected time given our radial velocities. No transit is evident.}
    \label{fig:tess-photometry}
\end{figure*}

TESS photometry for \starname\ was obtained in short-cadence 
(2-minute) integrations from July 2018 to August 2018 (during 
a total time-span of 27.9 days) in TESS Sector 1 using Camera 2. 
The TESS Science Processing Operations Center (SPOC) photometry was retrieved from the alerts webpage\footnote{\url{https://tess.mit.edu/alerts/}}, which 
provide either simple 
aperture photometry (\texttt{SAP\_FLUX}) or the 
systematics-corrected photometry (\texttt{PDCSAP\_FLUX}), a 
procedure performed by an adaptation of the Kepler Presearch 
Data Conditioning algorithm \citep[PDC, ][]{PDC1,PDC2,PDC3} to TESS. We use this latter photometry along with its 
provided errorbars (\texttt{PDCSAP\_FLUX\_ERR}) in the rest 
of this work; we refer to this photometry as the PDC 
photometry in what follows. Both, the SAP and PDC median-normalized photometry provided by the TESS alerts 
are presented in Figure \ref{fig:tess-photometry}. For the 
analysis that follows, we remove two portions of the data: 
the portion (in BJD - 2457000) between 1347.5 and 1349.3, 
which was obtained during 
a period of increased spacecraft pointing jitter \citep[see][]{pimen}, and the region 
after 1352, which shows an evident relatively short but significant decrease in flux which we found might give rise to biases in our 
analysis.

The TESS alerts diagnostics, generated using the tools outlined 
in \cite{DV}, \cite{SPOC} and \cite{Li:2019} which have been adapted to work 
with TESS data, present this system as having 
a 1-day transit signal present in the data, which we 
refer to as \starname.01. The transit signature of this planet passes all 
the Data Validation (DV) tests (e.g., comparison of even and odd transits to 
screen against eclipsing binaries, ghost diagnostic tests to help rule out 
scattered light or background eclipsing binaries, among others) but 
the difference image centroiding test, likely due to the star being 
slightly saturated. From a difference image analysis done within the DV, 
however, the transit source is coincident with the core of the stellar 
point spread function (PSF), so it is clear the transit events happen on 
the target and not in, e.g., nearby bright stars. In order to confirm this 
signal and search for additional ones in the photometry, we ran the Box Least-Squares algorithm \citep[BLS,][]{bls:2002} on the 
data using a \texttt{python} implementation of this algorithm by Daniel Foreman-Mackey, \texttt{bls.py}\footnote{\url{https://github.com/dfm/bls.py}}. Significances of the possible peaks were computed by running the algorithm on a mock dataset, which contained the same median flux as the TESS photometry, and to which we added white-gaussian noise whose standard-deviation was defined as the provided errorbars at each time-stamp. This procedure was ran 100 times, giving 100 BLS powers at each period, with which 
the mean BLS power and the corresponding standard 
deviation at each period was calculated. A peak in the BLS spectrum of the original data was then considered 
significant and was later inspected if it deviated by more than 5-sigma from 
this white-gaussian noise spectrum. We ran the 
BLS on the search of transits with periods between 0.1 
and 14 days (the latter chosen as around half the 
total time-span of the TESS observations; 5,000 periods 
were considered between those limits), searching 
for transits with durations between $q = 0.01$ and 
$q=0.09$ in phase-space.

Using the BLS on the PDC photometry, the largest peak in the 
BLS periodogram was located at around the same period as 
the one reported on the TESS alerts, i.e., at 
1.007 days, with a depth of around 200 ppm. The peak is highly significant --- greater than 100 standard deviations above the mean BLS power at this period. It is interesting to note that the transits of this planetary 
candidate are individually visible in the lightcurve of 
\starname\ presented in Figure \ref{fig:tess-photometry}, 
indicated in that figure by red lines. We removed the 
in-transit points corresponding to \starname.01 
and repeated the same procedure on the masked data. 
A couple of peaks emerge in the  
BLS periodogram just above our 5-sigma threshold, but when 
phasing the data at those periods, no evident transit signature 
emerges. In addition, those peaks are only at specific periods, 
and thus very narrow (one or two points) in the BLS periodogram. 
We thus conclude that no more significant transit-like signals are present in the BLS periodogram of our data. Possible additional signals in the photometry were also 
inspected using the Transiting Planet Search (TPS) within the SPOC Data Validation (DV) component, which as mentioned above 
has been recently adapted to work with TESS data 
\citep{DV,SPOC,Li:2019}. No additional transiting planets were found 
with those tools either.

It is important to note that the aperture used to obtain the 
TESS photometry encompasses about 3 TESS pixels in radius around 
the target, which amounts to an on-sky aperture of about 1 
arcminute which in turn could lead to the light of other stars 
to contaminate the aperture. {This could in turn give rise to} possible 
dilutions of the observed transits, which could lead to misdetermination of the transit parameters, and to possible 
false-positives, which could led us to believe this is an 
exoplanetary system when in reality the observed TESS transit 
events could be due to a blend with a nearby eclipsing binary. 
In particular, the TESS aperture includes light 
not only from \starname\ (which has a TESS magnitude of 
$7.358 \pm 0.018$), but {also from five nearby stars}: two faint stars, which we 
denote C1 and C2 in what follows, at about 30 arcseconds 
from the target detected by both 2MASS 
(2MASS IDs: 22360031-5952070 and 22355219-5952034, for 
C1 and C2, respectively) and Gaia \citep[Gaia Source IDs 6407428925971511808 and 6407428960331344512, 
for C1 and C2, respectively;][]{DR2:2018}), and three additional stars fainter 
than {C1 and C2} by Gaia (Gaia Source IDs: 6407428891610548736, 
6407428925970566272 and 6407434801486912768), the brightest 
of which has a Gaia magnitude of $G = 19.86$ --- 
implying a $\Delta G = 12.1$ with \starname. 

Assuming the 
magnitude difference in the TESS passband to be similar to the 
difference in the Gaia passband, the three faint stars detected 
by Gaia and not 2MASS are negligible sources of light in practice 
to the TESS aperture 
(if any of these were a totally eclipsing binary, for example, 
they would lead to transit depths of about 15 ppm; in terms of 
lightcurve dilution, they amount to less than $0.0015\%$ of 
the light in the aperture). For C1 and C2, using the relations 
in \cite{TIC}, their TESS magnitudes 
are $T_\textnormal{C1} = 16.862 \pm 0.025$ and 
$T_\textnormal{C2} = 16.417 \pm 0.023$ respectively 
(calculated using the 2MASS $J$ and Gaia $G$ magnitudes of these stars, which are the magnitudes that have the smaller 
errorbars, and propagating the errors on the relations of 
\cite{TIC} in quadrature to the photometric errors). This implies 
a magnitude difference with TOI-141 in the TESS passband of 
$9.504 \pm 0.031$ 
and $9.059 \pm 0.029$ for C1 and C2 respectively, thus amounting 
for $0.041$\% of the light 
in the TESS aperture. If any of those stars were to produce the 
observed transits in the TESS photometry of \starname.01, they would have to be variable objects producing periodic 1-day 
dimmings of at least $80\%$ of their light. We explore 
this possibility with follow-up lightcurves in the next sub-section.

\subsection{Photometric follow-up}
Photometric follow-up was performed as part of the TESS Follow-up 
Program (TFOP) SG1 Group. We used the {\tt TESS Transit Finder}, which is a customized version of the {\tt Tapir} software package \citep{Jensen:2013}, to schedule photometric time-series follow-up observations. Observations 
of \starname.01 were obtained on September 11, 2018, 
using the CDK700 27-inch telescope at Mount Kent Observatory 
(MKO). The observations were made in $r'$ using 128 second 
``deep'' exposures, effectively saturating \starname\ but gathering 
enough photons to provide precise photometry for the fainter 
companion stars in order to rule-out false-positive scenarios. 
The observations covered around 3 hours, and effectively 
covered the predicted ingress and egress events. We used AstroImageJ \citep{Collins:2018} to calibrate the data and extract the differential aperture photometry of the target and nearby stars. All stars within $2\arcmin$ turned out to have a constant brightness to within 10\%. Dimmings at the 80\% or larger 
for C1 and C2 can be confidently ruled out by these 
observations; however, C1 showed a 70\% \textit{rise} in the photometry around the expected mid-transit time of \starname.01, 
which was due to 
an instrumental effect: due to the rotation of the field, some 
of the diffraction spikes of \starname\ fell on the aperture 
used to extract the photometry of C1 at these times generating 
this increase in the relative flux of this object. It is important to stress here that although mid-transit was lost due to this effect, the lightcurve before this event showed no large variations as the ones expected from an eclipsing binary causing the TESS transits (the precise transit ephemerides for this 
system ensure we should have caught at least an ingress 
event if this was indeed an eclipsing binary).

The observations presented above thus rule out any possible 
near eclipsing binary as being the responsible for the transit 
events observed in the TESS lightcurve.

\subsection{Speckle imaging}
\begin{figure}
   \includegraphics[height=0.85\columnwidth]{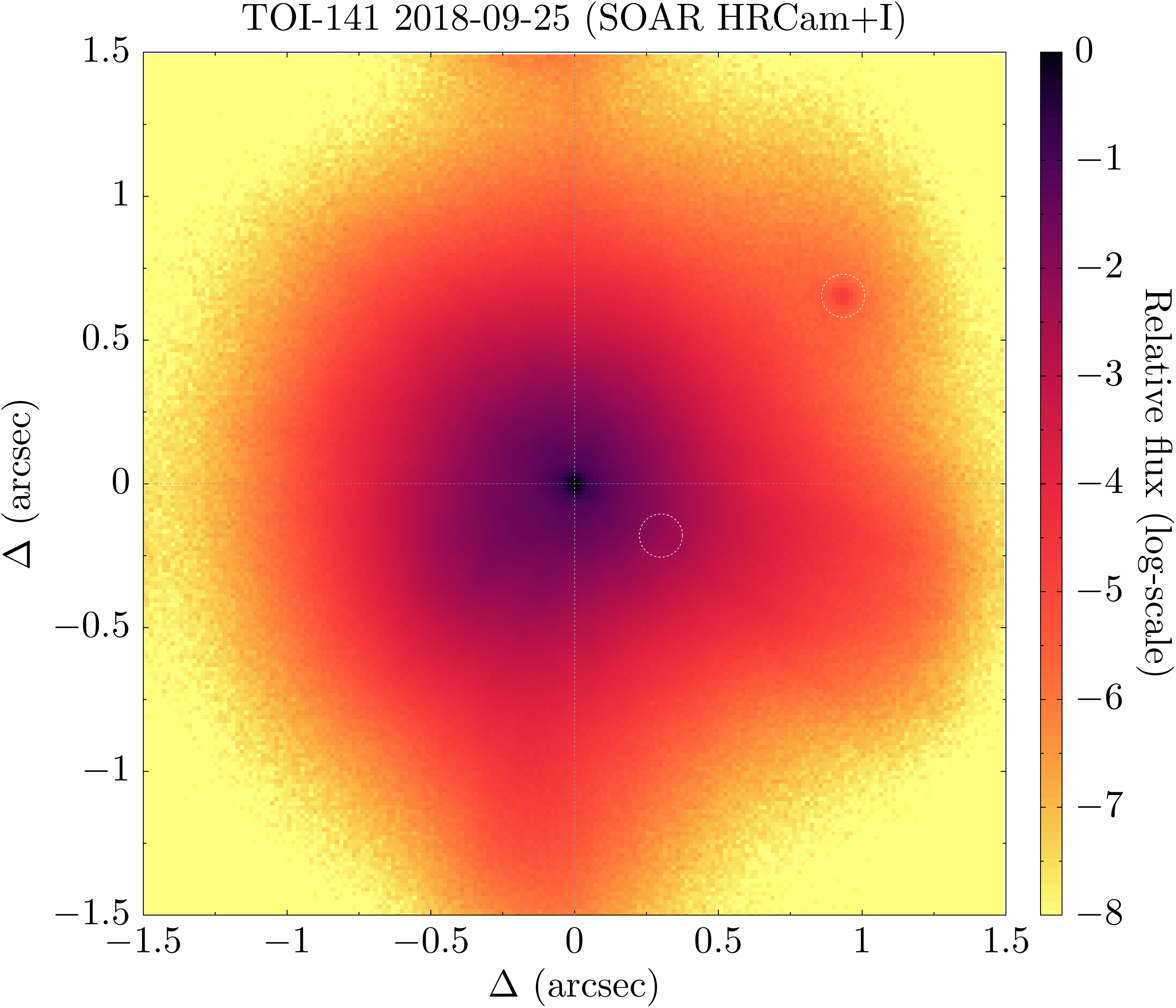}
    \caption{``Lucky" image obtained with the HRCam at the 4.1m 
    SOAR telescope in the I band for \starname. The distant 
    companion at 
    1.19'' (indicated by a white circle in the figure) is 
    evident from the image, whereas the closer 0.4'' companion 
    (also indicated) is not; this was detected via speckle ACF (see text).}
    \label{fig:imaging}
\end{figure}
Speckle imaging for \starname\ was obtained on September 24, 
2018, using the High-Resolution Camera (HRCam) at the 4.1m Southern 
Astrophysical Research (SOAR) telescope located in Cerro Pach\'on, Chile, 
in the $I$-band; the co-added images are presented in Figure \ref{fig:imaging}. The instrument and 
the corresponding 
analysis and reductions of data obtained with it is detailed 
in \cite{SOAR:2018}. These observations, and the subsequent analysis of the auto-correlation function (ACF) of the image, which provides better dynamic range than working on the images directly \cite[see][for details]{SOAR:2018}, 
reveal 2 companions to \starname: one at a separation of 1.19'' 
from \starname\ at an angle of 305 degrees, and another at 
a separation of 0.4'' from the target at 239 degrees, 
with magnitude differences of $\Delta I = 5.4$ and 
$\Delta I = 4.9$ respectively. As will be shown in Section \ref{sec:analysis}, given the observed radial-velocity 
variations in phase with the transit ephemerides observed by 
TESS --- and given these companion stars are too faint to 
produce any measurable signal in our radial-velocity measurements 
--- it is very unlikely the companion stars revealed by 
these speckle imaging observations are the ones producing 
the transit events. These stars, however, could be important to 
constrain the possible transit dilutions they imply 
for our target. However, given these objects are not 
detected in Gaia DR2 \citep{DR2:2018}, and that only 
one-band observations are available, we cannot calculate 
either if they are physically bound nor their predicted 
TESS magnitudes in order to calculate the dilution these 
stars would imply in the TESS bandpass. If we assume the 
delta-magnitudes in $I$ band are similar to the TESS magnitudes, 
then these stars would account for about 1.8\% of the total flux 
in the TESS aperture. For a 200 ppm transit depth as the 
one detected by the TESS photometry for \starname.01, this 
would imply a dilution of about the same percent of this 
transit depth (i.e., a depth about 4 ppm smaller) --- this is 
well below the error on the transit depth, which as it will be 
shown in Section \ref{sec:analysis}, is on the order of 12 ppm.

It is important to notice that, because \starname\ is a relatively close 
system ($48$ pc --- see Section \ref{sec:analysis}), monitoring the 
system via high angular resolution in the future (e.g., a few years) might 
reveal if these companions detected with our observations are physically 
associated or not to \starname. We encourage future observations in order 
to determine if this is the case.

\subsection{FEROS radial velocities}

High precision radial velocities were obtained for 
\starname\ with the 
Fiber-fed Extended Range Optical Spectrograph 
\citep[FEROS, ][]{feros} mounted at the 2.2m MPG telescope at 
La Silla Observatory between September and October 2018 by the 
Chile-MPIA group. A total of 175 RV measurements were obtained {with a 
simultaneous ThAr calibration using 200s exposures. The RVs were extracted from 
the spectra using the customized} CERES pipeline \citep{ceres}, {which performs 
all the process of extraction from basic bias, dark and flat-field corrections (including 
scattered light) to order tracing, wavelength 
calibration and cross-correlation matching of the spectrum with templates to 
obtain the RVs from the spectra}. Although 
based on standard stars the precision that the CERES pipeline 
obtains with FEROS is 7 m/s, we found that with some modifications 
to the standard acquisition of FEROS frames one can achieve 
3 m/s precision for $V = 8$ stars: simply by turning the ThAr 
lamp around 20 minutes before it is used, and taking a long 
series of ThAr calibration images to select the best one as
reference greatly improves the precision one can achieve with 
FEROS using CERES. We followed these procedures for the obtention 
of the RVs of \starname\ and imposed this 3 m/s noise floor 
to the star based on the monitoring of standard stars.

The FEROS observations showed radial velocities in phase 
with the transit ephemerides of \starname.01, showing an 
amplitude of about 5 m/s. In addition, they also showed an 
evident extra sinusoidal variation at a period of about 
$4.78$-days. These signals will be analyzed in detail in 
Section \ref{sec:analysis}. The data is presented in Table \ref{rvs}.

\begin{table*}
 \centering 
 \begin{center}
\caption{Radial-velocity measurements obtained for \starname.}
 \label{rvs}
 \begin{threeparttable}
  \centering
  \begin{tabular}{ lllll }
   \hline
   \hline
     Time (BJD) & Radial-velocity (m/s) & Error (m/s) & Exposure time (s) & Instrument \\
   \hline
2458409.7085776 & 36164.16 & 0.56 & 900 & HARPS\\
2458412.4991098 & 36146.57 & 0.39 & 900 & HARPS\\
2458412.5935471 & 36149.97 & 0.46  & 900 & HARPS\\
2458412.6969091 & 36153.13 & 0.55 & 900 & HARPS\\
\vdots & \vdots & \vdots & \vdots\\
   \hline

   \end{tabular}
      \textit{Note}. This table will be available in machine-readable form in the 
      online journal. A portion is shown here for guidance regarding its form and content.
  \end{threeparttable}
 \end{center}
 \end{table*}

\subsection{HARPS radial velocities} 
High precision radial velocities were also obtained with the 
High Accuracy Radial velocity Planet Searcher (HARPS) mounted 
at the ESO La Silla 3.6m telescope \citep{harps}. These data 
were obtained by three groups: the Chile-MPIA group (14 
measurements in September 2018), the NCORES 
group (14 measurements in October 2018) and the U. de Chile 
group (19 measurements between October and November 2018){, all of 
which were obtained using simultaneous ThAr calibration lamps}. 
In total, 47 measurements were obtained for \starname\ between 
September 2018 and November 2018. {The conditions during the September 2018 run 
were sub-optimal, which in turn led us to use longer exposure times on those nights of 900 seconds. 
Conditions were photometric for the rest of the observing runs, and so 300s exposures were used 
in those nights to gather spectroscopic measurements for \starname}. The radial velocities were 
obtained with both the CERES pipeline \citep{ceres} and the 
HARPS DRS pipeline. Although both gave consistent results, the 
CERES pipeline results have much larger long-term errors as monitored 
by standard stars than the quoted attainable precision by the DRS pipeline. 
Because of this, we decided to use the DRS instead of the CERES results in 
this work. The RV precision of those datapoints varied with the exposure times --- 
0.5 m/s for 900 second exposures and 2 m/s for 300 second exposures. The data is presented in Table \ref{rvs}.

\subsection{CORALIE radial velocities}
\label{data:coralie}
Additional data were obtained with the CORALIE instrument, mounted on the 1.2m Euler Telescope at the La 
Silla Observatory \citep{coralie} both prior to the TESS observations and after the TESS observations. A first set of data, here denoted CORALIE07, were taken between August 2008 and October 2013 (7 radial-velocity measurements) 
and a second set of data, here denoted CORALIE14, were taken between August 2016 and September 2018 (8 radial-velocity 
measurements). From this latter set, 6 datapoints were taken after the TESS alerts were released. These datapoints have precisions between 3-4 m/s, and the radial velocities were analyzed with the official CORALIE pipeline.  Two extra datapoints to the just mentioned ones were obtained in 
July 1990 and August 1994 by CORAVEL \citep{coravel}, and another set of 12 radial-velocity datapoints were taken between September 2001 and September 2006 with CORALIE; however, we do not use 
those measurements in this work as they show errors in excess of the precision 
needed to constrain the masses of the exoplanets presented in this work. 
In total, thus, in this work we use 15 radial-velocity measurements from CORALIE.

It is important to note here that the CORALIE instrument was upgraded 
in November 2014 \citep[see, e.g.][]{maxted:2016}. This means that the zero-point offset between the CORALIE07 and CORALIE14 radial velocities is different. Because 
of this reason, here we treat each as an independant dataset, meaning 
that in the analysis to be described in Section \ref{sec:jointanalysis}, we 
consider different systemic velocities and jitters for each of those datasets. The data is presented in Table \ref{rvs}.

\section{Analysis}
\label{sec:analysis}
\subsection{Stellar properties}
\label{sec:sprops}
We followed the iterative procedure described in \citet{k2-232} and \citet{k2-161}
to determine the physical parameters of \starname.
First we used the co-added HARPS spectra to compute the atmospheric
parameters of \starname\ by using the \texttt{ZASPE} code \citep{zaspe},
which compares the observed spectrum with synthetic ones in the spectral
regions that are most sensitive to changes in the atmospheric parameters.
Then we combined the GAIA DR2 parallax and the available photometry 
to compute the stellar radius and extinction using an MCMC
code\footnote{\url{https://github.com/rabrahm/rstar}}. Finally,
we determined the stellar mass and age by searching for the Yonsei-Yale
evolutionary model \citep{YY}  that matched the observed stellar radius
and spectroscopic effective temperature through another MCMC code\footnote{\url{https://github.com/rabrahm/isoAR}}.
With the derived stellar mass
and radius we computed a new value for the log(g) which is held fixed
in a new \texttt{ZASPE} iteration, followed by the same steps that were
just described. The final stellar parameters obtained for \starname\ are
presented in Table \ref{stellar}.

\subsection{Stellar abundances}
\label{sec:sab}
Stellar abundances are important to constrain possible interior composition models of exoplanets orbiting host stars, as they can give prior information to be used by structure modeling in order to constrain the composition of an exoplanet (see Section \ref{discussion:composition}). Because of this, we extracted abundances from the HARPS (co-added) spectra of important refractory and volatile elements that could aid as priors in such an analysis. We use a standard LTE analysis with the 2017 version of MOOG \citep{moog} and Kurucz ATLAS9 model atmospheres \citep{atlas9}, measuring the equivalent widths of Si, Ni, Mg, and C lines.
The C abundance is based on two unblended lines at $5052.2$ and $5380.3 \textnormal{\AA}$ with atomic parameters from \citet{delgadomena2010c}. For Si, Ni, and Mg, the line list of \citet{neves2009linelist} is used instead.
We found abundances $A(\textnormal{X}) = \log_{10} N(\textnormal{X})/N(\textnormal{H}) + 12$, where $N(\textnormal{X})/N(\textnormal{H})$ is the atomic ratio between element X and hydrogen (H), where $N(\textnormal{H})$ is normalized to $10^{12}$ hydrogen
atoms (i.e., $N(\textnormal{H})=12$) for Si, Ni, Mg and C --- those are presented in Table \ref{stellar}.
The errors reported consider the line-by-line scatter added in quadrature with errors produced by uncertainties on stellar parameters $T_\textnormal{eff}$, [Fe/H], and $\log g_*$.

\begin{table}
 \centering 
 \begin{center}
\caption{Stellar parameters of \starname.}
 \label{stellar}
 \begin{threeparttable}
  \centering
  \begin{tabular}{ lcr }
   \hline
   \hline
     Parameter &  Value & Source \\
   \hline
Identifying Information\\
~~~TIC ID & 403224672 & TIC\\
~~~GAIA ID & 6407428994690988928 & Gaia DR2\\
~~~2MASS ID & 22355630-5951522 & 2MASS\\
~~~R.A. (J2015.5, h:m:s) & 22$^h$35$^m$56.09$s$ & Gaia DR2\\
~~~DEC (J2015.5, d:m:s) & -59$^o$51$'$53.38$''$ & Gaia DR2\\
Spectroscopic properties\\
~~~$T_\textnormal{eff}$ (K) & $5978 \pm 50$& ZASPE\\
~~~Spectral Type & G & ZASPE\\
~~~[Fe/H] (dex) & $-0.04\pm 0.03$ & ZASPE\\
~~~$\log g_*$ (cgs)& $4.3827^{+0.0095}_{-0.0097}$ & ZASPE\\
~~~$v\sin(i_*)$ (km/s)& $3.0\pm 0.2$ & ZASPE\\
~~~$A(\textnormal{Si})$ (dex) & $7.48 \pm 0.09$ & MOOG\\
~~~$A(\textnormal{Ni})$ (dex) & $6.19 \pm 0.11$ & MOOG\\
~~~$A(\textnormal{Mg})$ (dex) & $7.51 \pm 0.06$ & MOOG\\
~~~$A(\textnormal{C})$ (dex) & $8.31 \pm 0.13$ & MOOG\\
Photometric properties\\\
~~~$T$ (mag)& $7.358 \pm 0.018$ & TESS\\
~~~$B$ (mag)& $8.4720\pm 0.0020$ & APASS\\
~~~$V$ (mag)& $7.9960\pm0.0020$ & APASS\\
~~~$r'$ (mag)& $7.8500\pm0.0010$ & APASS\\
~~~$i'$ (mag)& $7.7130\pm0.0020$ & APASS\\
~~~$z'$ (mag)& $7.4690\pm0.0020$ & APASS\\
~~~$J$ (mag)& $6.806\pm0.015$ & 2MASS\\
~~~$H$ (mag)& $6.501\pm0.031$ & 2MASS\\
~~~$Ks$ (mag)& $6.419\pm0.019$ & 2MASS\\
Derived properties\\
\vspace{0.1cm}
~~~$M_*$ ($M_\odot$)& $1.068^{+0.020}_{-0.018}$ & YY$^{*}$\\
~~~$R_*$ ($R_\odot$)& $1.1011^{+0.0080}_{-0.0075}$ & Gaia DR2$^{*}$\\
~~~$L_*$ ($L_\odot$)& $1.376^{+0.045}_-{0.049}$ & YY$^{*}$\\
~~~$M_V$ & $4.462^{+0.044}_{-0.042}$ &  YY$^{*}$\\
~~~Age (Gyr)& $3.80^{+0.66}_{-0.79}$ & YY$^{*}$\\
~~~Distance (pc)& $47.97 \pm 0.14 $ & Gaia DR2+YY$^{*}$\\
~~~$\rho_*$ (kg m$^{-3}$)& $1127\pm 33$ & YY$^{*}$\\
   \hline

   \end{tabular}
      \textit{Note}. Logarithms given in base 10.\\
      *: Using stellar parameters obtained from ZASPE.
  \end{threeparttable}
 \end{center}
 \end{table}

\subsection{Joint analysis}
\label{sec:jointanalysis}
The joint analysis of the photometry and radial velocities 
is performed here using a new code introduced in 
\cite{juliet}, \codename, which is available via GitHub\footnote{\url{https://github.com/nespinoza/juliet}}. For 
our analysis in this work, \codename\ uses \texttt{batman} \citep{batman:2015} to model the transit lightcurves and 
\texttt{radvel} \citep{radvel:2018} to model the radial velocities. 
\codename\ allows for a 
variety of parametrizations, and in particular allows us to 
incorporate Gaussian Processes (GPs) via the \texttt{george} 
\citep{george} and \texttt{celerite} \citep{celerite} packages, 
which are implemented within \codename\ for modelling 
underlying systematic and/or astrophysical signals present 
either in the radial velocities, the photometry or both, and to 
easily incorporate those into our modelling scheme. One of the key 
features of \codename\ is its ability 
to perform model comparison, as nested sampling algorithms are 
used to compute posterior samples and, in particular, model 
evidences, $Z_i$, for a model $M_i$ given the data, 
$\mathcal{D}$, i.e., $Z_i = p(\mathcal{D}|M_i)$. In this 
work within \codename\ we make use of MultiNest \citep{MultiNest} 
via the PyMultiNest wrapper \citep{PyMultiNest} to explore the 
parameter space and perform model evidence calculations. This evidence 
estimation in turn 
allows to compute the probability of the model given the data, $p(M_i|\mathcal{D}) = p(M_i)p(\mathcal{D}|M_i)$ given a prior probability for model $M_i$, $p(M_i)$. Here, unless otherwise stated, 
we assume all models are a-priori equiprobable and thus 
compare model evidences directly between models as 
in this case the posterior odds are simply $p(M_i|\mathcal{D})/p(M_j|\mathcal{D}) = Z_i/Z_j$. For 
ease of comparison, we here compare models using the 
difference of the (natural) log-evidences, $\Delta \ln Z_{i,j} = \ln Z_i/Z_j$. We have taken care to repeat the model 
evidence calculations several times in order to account 
for the miscalculation of errors on evidences known to 
happen in nested sampling algorithms \citep[see][]{Nelson:2018}; however, we note that in our case, given the large amount of data (especially given we have strong constraints on the 
ephemerides of at least one planet in this work from 
transit photometry), the empirically determined errors on the evidences (calculated by running each model run five times) are 
always $\ln Z < 1$ --- typically on the order of $0.1$.

\begin{table*}
    \centering
    \caption{Priors used in our joint analysis of the \starname\ system using \codename\ for the analysis of \planetnameb\ and \planetnamec. Our stellar density prior is the one derived in Section \ref{sec:sprops}. Here $p=R_p/R_*$ and $b=(a/R_*)\cos(i_p)$, where $R_p$ is the planetary 
    radius, $R_*$ the stellar radius, $a$ the semi-major axis of the 
    orbit and $i_p$ the inclination of the planetary orbit with respect 
    to the plane of the sky. $e$ and $\omega$ are the eccentricity and 
    argument of periastron of the orbits. $\mathcal{N}(\mu,\sigma^2)$ represents a normal distribution of mean $\mu$ and variance $\sigma^2$. $\mathcal{U}(a,b)$ represents a uniform distribution between $a$ and $b$. $\mathcal{J}(a,b)$ represents a Jeffrey's 
    prior (i.e., a log-uniform distribution) between $a$ and $b$.}  
    \label{tab:priors}
    \begin{tabular}{lccl} 
        \hline
        \hline
        Parameter name & Prior & Units & Description \\
                \hline
                \hline
        Parameters for \starname & & \\
        ~~~$\rho_*$ &$\mathcal{N}(1127,33^2)$ & kg/m$^3$ & Stellar density of \starname. \\
        \hline
        Parameters for \planetnameb & & \\
        ~~~$P_b$ &$\mathcal{N}(1.0079,0.0100^2)$ & days & Period of \planetnameb. \\
        ~~~$t_{0,b}$ &$\mathcal{N}(2458379.9647,0.0100^2)$ & days & Time of transit-center for \planetnameb. \\
        ~~~$r_{1,b}$ &$\mathcal{U}(0,1)$ & --- & Parametrization$^{1}$ of \cite{Espinoza:2018} for $p$ and $b$ for \planetnameb. \\
        ~~~$r_{2,b}$ &$\mathcal{U}(0,1)$ & --- & Parametrization$^{1}$ of \cite{Espinoza:2018} for $p$ and $b$ for \planetnameb. \\
        ~~~$K_{b}$ &$\mathcal{U}(0,100)$ & m/s & Radial-velocity semi-amplitude for \planetnameb. \\
        ~~~$\mathcal{S}_{1,b} = \sqrt{e_b}\sin \omega_b$ &$\mathcal{U}(-1,1)$ & --- & Parametrization$^{2}$ for $e$ and $\omega$ for \planetnameb. \\
        ~~~$\mathcal{S}_{2,b} = \sqrt{e_b}\cos \omega_b$ &$\mathcal{U}(-1,1)$ & --- & Parametrization$^{2}$ for $e$ and $\omega$ for \planetnameb. \\
        \hline
        Parameters for \planetnamec & & \\
        ~~~$P_c$ &$\mathcal{N}(4.75,1.00^2)$ & days & Period of \planetnamec. \\
        ~~~$t_{0,c}$ &$\mathcal{N}(2458397.00,1.00^2)$ & days & Time of transit-center for \planetnamec. \\
        ~~~$K_{c}$ &$\mathcal{U}(0,100)$ & m/s & Radial-velocity semi-amplitude for \planetnamec. \\
        ~~~$\mathcal{S}_{1,c} = \sqrt{e_c}\sin \omega_c$ &$\mathcal{U}(-1,1)$ & --- & Parametrization$^{2}$ for $e$ and $\omega$ \planetnamec. \\
        ~~~$\mathcal{S}_{2,c} = \sqrt{e_c}\cos \omega_c$ &$\mathcal{U}(-1,1)$ & --- & Parametrization$^{2}$ for $e$ and $\omega$ \planetnamec. \\
        \hline
        Parameters for TESS photometry & & \\
        ~~~$D_{\textnormal{TESS}}$ & 1 (fixed) & --- & Dilution factor for TESS. \\
        ~~~$M_{\textnormal{TESS}}$ &$\mathcal{N}(0,0.1^2)$ & relative flux & Relative flux offset for TESS. \\
        ~~~$\sigma_{w,\textnormal{TESS}}$ &$\mathcal{J}(0.1,5000^2)$ & relative flux (ppm) & Extra jitter term for TESS lightcurve. \\
        ~~~$q_{1,\textnormal{TESS}}$ &$\mathcal{U}(0,1)$ & --- & Quadratic limb-darkening parametrization$^{3}$ \citep{kipping:2013}. \\
        ~~~$q_{2,\textnormal{TESS}}$ &$\mathcal{U}(0,1)$ & --- & Quadratic limb-darkening parametrization$^{3}$ \citep{kipping:2013}. \\
        \hline
        RV parameters & & \\
        ~~~$\mu_{\textnormal{FEROS}}$ &$\mathcal{N}(36140,30^2)$ & m/s & Systemic velocity for FEROS. \\
        ~~~$\sigma_{w,\textnormal{FEROS}}$ &$\mathcal{J}(0.01,30^2)$ & m/s & Extra jitter term for FEROS. \\
        ~~~$\mu_{\textnormal{HARPS}}$ &$\mathcal{N}(36162,30^2)$ & m/s & Systemic velocity for HARPS. \\
        ~~~$\sigma_{w,\textnormal{HARPS}}$ &$\mathcal{J}(0.01,30^2)$ & m/s & Extra jitter term for HARPS. \\
        ~~~$\mu_{\textnormal{CORALIE07}}$ &$\mathcal{N}(36088,30^2)$ & m/s & Systemic velocity for CORALIE07$^{4}$. \\
        ~~~$\sigma_{w,\textnormal{CORALIE07}}$ &$\mathcal{J}(0.01,30^2)$ & m/s & Extra jitter term for CORALIE07$^{4}$. \\
        ~~~$\mu_{\textnormal{CORALIE14}}$ &$\mathcal{N}(36135,30^2)$ & m/s & Systemic velocity for CORALIE14$^{4}$. \\
        ~~~$\sigma_{w,\textnormal{CORALIE14}}$ &$\mathcal{J}(0.01,30^2)$ & m/s & Extra jitter term for CORALIE14$^{4}$\\
        \hline
        \hline
    \end{tabular}
    \begin{tablenotes}
      \small
      \item $^1$ To perform the transformation between the $(r_1,r_2)$ plane and the $(b,p)$ plane, we performed the 
      transformations outlined in \cite{Espinoza:2018}, which depend on $r_{1}$ and $r_{2}$, and a set of limits for the minimum and maximum $p$, $p_l$ and $p_u$, to consider: if $r_{1}>A_r = (p_u-p_l)/(2 + p_l + p_u)$, then $(b,p) = ([1+p_l][1+(r_{1}-1)/(1-A_r)], (1-r_{2})p_l + r_{2}p_u)$. If 
      $r_{1}\leq A_r$, then $(b,p) = ([1+p_l] + \sqrt{r_{1}/A_r}r_{2}(p_u-p_l), p_u + (p_l-p_u)\sqrt{r_{1}/A_r}[1-r_{2}])$. In this work, we set $p_l=0$ and $p_u=1$.
      \item $^2$ We ensure in each sampling iteration that $e=\mathcal{S}^2_1+\mathcal{S}^2_2 \leq 1$.
      \item $^3$ To transform from the $(q_1,q_2)$ plane to the plane of 
      the quadratic limb-darkening coefficients, $(u_1,u_2)$, we use 
      the transformations outlined in \cite{kipping:2013} for this 
      law $u_1 = 2\sqrt{q_1}q_2$ and $u_2=\sqrt{q_1}(1-2q_2)$.
      \item $^{4}$ CORALIE07 corresponds to data taken between the 2007 and 2014 upgrade and CORALIE14 corresponds to data taken 
      after the 2014 upgrade (see Section \ref{data:coralie}). 
    \end{tablenotes}
\end{table*}

\subsubsection{Photometry-only analysis}
For the analysis of the \starname\ system, we first performed 
a photometry-only analysis with \codename\ in order to find 
constraints on the time of transit center and period 
of the orbit of \starname.01\ using the priors defined 
for the photometric elements in Table \ref{tab:priors}, which 
were based on our BLS search and the TESS alerts best period 
for this candidate. We consider the possibility that the 
TESS photometry might need a GP to account for any 
residual time-correlated noise in the lightcurve, and 
for this we fitted both a transit model plus an 
exponential-squared GP and a transit model assuming a white-noise 
model only. We found that both models were indistinguishable 
from one another based on their model evidences 
($\Delta \ln Z < 1$), and thus decided to use the simpler 
model (i.e., a no-GP, white-noise model) when analyzing the 
PDC photometry. We note that for the white-noise model we add an extra photometric jitter term 
in quadrature to the reported uncertainties in order to account 
for miscalculations of the photometric uncertainties or any 
residual astrophysical signal not captured by our modelling. 

In addition to this fit, we also tried a fit assuming there is 
an additional transiting planet in the system to \starname.01, 
with the same photometric priors as the ones used for 
this candidate presented in Table \ref{tab:priors}, 
except for the period and time of transit center; the first 
was left to freely vary between 0.1 and 14 days (for the 
same reason this were the trial periods in our BLS analysis 
in Section \ref{sec:data}), whereas the second was left to 
vary from the time of the start of the observations to 15 
days later. We found no evidence on the data for 
additional transiting planets ($\ln Z > 100$ in favor of 
the 1-planet model) in agreement with our results from the 
BLS search in Section \ref{sec:data}.

\subsubsection{RV-only analysis}
We ran a \codename\ run on the radial velocities independently 
in order to see if we were able to find evidence for 
planets in the radial-velocity 
dataset alone. For this, we ran three models: (1) no 
planet (i.e., variation in the data solely explained 
by the jitters of the data, which were let to 
float as free parameters), (2) one planet in the RVs, (3) 
two planets in the RVs. We modelled the planetary signals using 
simple Keplerians assuming circular orbits with the same priors 
as the radial-velocity elements in Table \ref{tab:priors}. However, 
for this excercise we gave wide log-uniform priors for the period 
from 0.1 to 30 days for both planets (with the constraint that 
one planetary period is always larger than the other in order to 
avoid multiple modes for exchangable periods) and the times 
of transit center set with uniform priors between the start of 
the observations and 30 days later\footnote{In practice, 
this gave rise to many local minima corresponding to integer times 
the period along the observations but this is not a problem 
for the nested sampling algorihtms used by \codename --- see \cite{juliet} for details on this point.}. The limit 
of 30 days was set as our most constraining RV datasets (the 
FEROS and HARPS datasets) are only $\sim 60$ days in total 
duration, and as such periods up to half this baseline 
are reasonable to search in the dataset.

The resulting evidences for the models strongly favor the 1 and 
2-planet models in the data over the no-planet model. 
The 1-planet model converges to a posterior period of 
$4.75 \pm 0.01$ days, and it has a log-evidence 56 times 
larger than the no-planet model, i.e., the 1-planet model 
is 24 orders of magnitude more likely than the null model. In 
turn, the two-planet model converges to both a period of 
$1.00940 \pm 0.00036$ days for one of the planets and 
of $4.7604\pm 0.0028$ days for the other --- this model in 
turn has a log-evidence 52 times larger than the 1-planet 
model, and 108 times larger than the no-planet model. We note 
how in this 2-planet model the smallest period is consistent 
with the period of the transit events observed by TESS, albeit 
with a small offset, most likely due to the sampling of the 
data \citep[i.e., given a signal with a period equal to that 
of the transit ephemerides in our data, this offset is 
expected given the alias of 1-day the window function 
imprints on our radial-velocity measurements; see][for details, 
and our discussion below]{DF:2010}. This acts as an independant 
confirmation of the transit signal observed in the TESS photometry 
--- we consider these observations thus confidently 
confirm the transit signatures observed by TESS as a 
bona-fide exoplanetary signal, to which we refer to 
as \planetnameb\ in what follows.

\begin{figure*}
   \includegraphics[height=1.5\columnwidth]{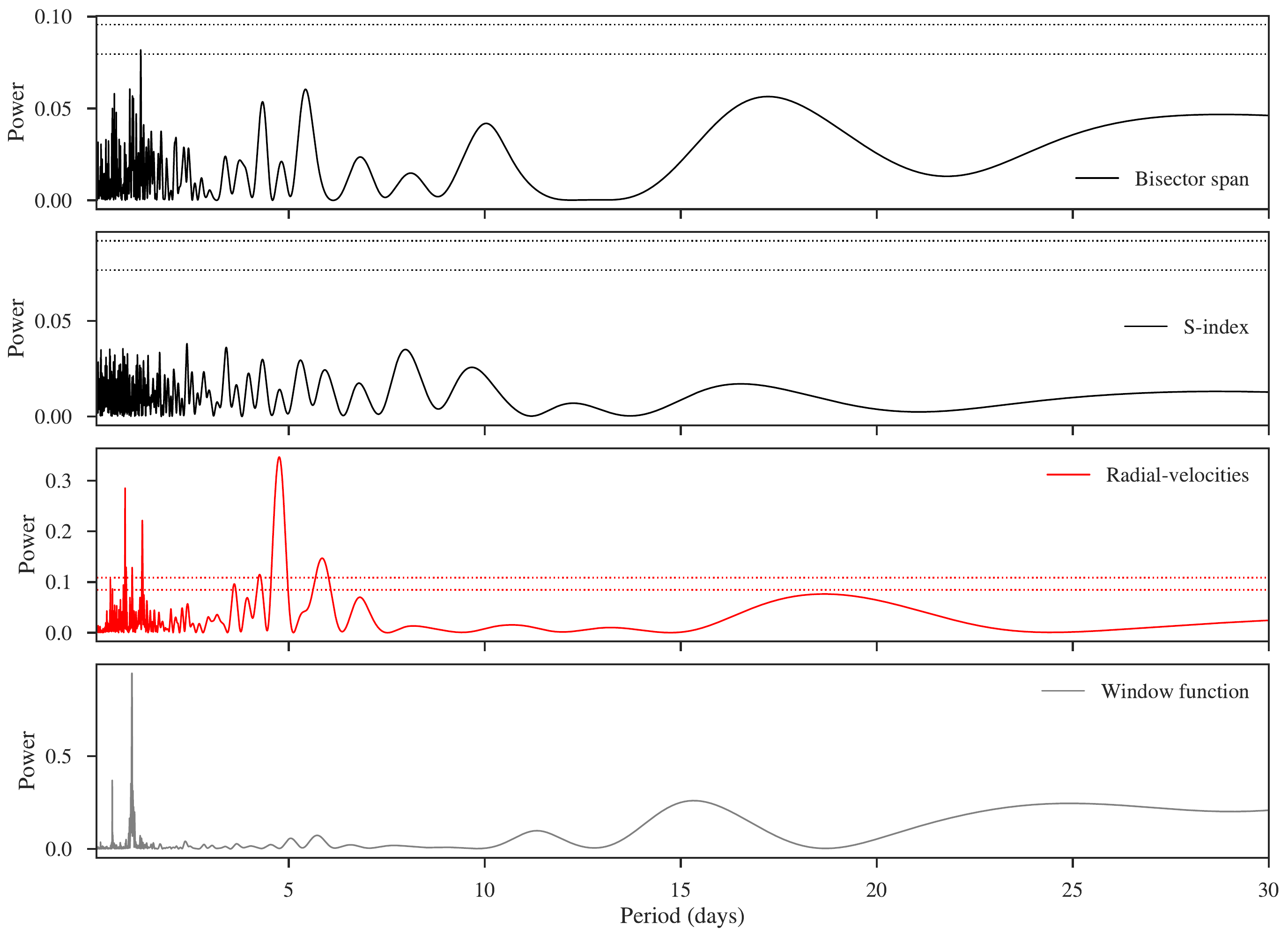}
    \caption{Generalized Periodogram for the bisector span (top), 
    S-index (second panel), the radial velocities (third panel) and 
    the window function (bottom). Dotted lines in each panel denote the 
    1\% and 5\% false-alarm probabilities. Periodograms calculated with 
    the Generalized Lomb Scargle periodogram \citep{GLS} and with 
    false-alarm probabilities calculated via bootstrapping with \texttt{astroML} \protect\citep{astroML}.}
    \label{fig:periodogram}
\end{figure*}

The $4.8$-day signal, although 
well-fitted with a Keplerian, could also be caused by stellar 
activity and not by the reflex motion of a planet around the 
star. We anticipate that this is not very likely, 
as the star's chromospheric emission as measured by the 
$\log R_\textnormal{HK}'$ has been actually measured before 
our observations to be quite low 
\citep[$-4.90\pm 0.05$;][]{henry:96}, which combined with 
its $B-V=0.62$ color, would imply it resides in the region 
where inactive stars reside in the $B-V$/$\log R_\textnormal{HK}'$ 
diagram. On top of this, assuming the stellar axis is 
aligned with the plane of the sky, we can derive a 
rotation period of the (equator of the) star of 
$18.58\pm 1.28$ days from the stellar radius and 
the $v\sin i_*$ value presented in Table \ref{stellar}, 
which is much too large to explain the evident 4.8-day 
variations observed in our radial velocities. Indeed, the periodogram of monitored external variables, such 
as Mount Wilson's S-index shows no clear peak around 
the periods of interest, and the same results are obtained 
for the bisector span (Figure \ref{fig:periodogram}). 
We nonetheless consider this possibility in the 
next sub-section when we perform the joint photometric 
and RV analysis.

\subsubsection{Photometric and RV analysis}
With the above defined information, we performed a 
joint analysis of the photometry and radial-velocity 
of \starname\ using \codename, which we 
use to jointly constrain all the parameters of the orbits 
of both \planetnameb\ and the possible 4.75-day planetary signal 
in the \starname\ system. We use normal priors for the 
periods and time of transit centers of those signals, with mean 
values taken from our photometry and radial-velocity only 
analyses, and with standard-deviations enlarged by a factor of 
a thousand with respect to those found in those analyses. All 
the other parameters are left to explore the whole parameter 
space of physically plausible ranges.

{In order to study the nature of the 4.8-day signal found in 
our radial-velocity only analysis and any possible additional signals in the radial velocities, 
we performed two groups of joint analyses in which we explored (1) how strongly this 
extra signal is supported by the data, (2) what the nature of this extra signal is (i.e., 
planetary or stellar activity) and (3) if there is any evidence for additional 
signals on top of this extra signal in the radial velocities. To explore (1) and (2) we considered 
two possible models for this extra signal: a Keplerian or a GP. For the Keplerian model, we used the 
priors presented in Table \ref{tab:priorsGP}. For the GP model, we used 
the same priors but instead of adding the parameters corresponding to planet c, we used a GP to account 
for the extra signal with three different possible kernels. The first was a squared-exponential kernel using either time, 
S-index or bisector spans as inputs, i.e., a kernel of the form}
\begin{equation*}
k_{i,j}(\tau) = \sigma_{\textnormal{GP}}^2 \exp \left(-\alpha_{\textnormal{GP}} \tau^2\right),
\end{equation*}
{where $\tau = x_{i} - x_{j}$, is the lag between the mentioned 
state-variables (which were fed normalized --- i.e., they were 
mean-substracted and divided by their standard-deviations), $\sigma_{\textnormal{GP}}$ is the amplitude of this 
GP component and $\alpha_{\textnormal{GP}}$ is the inverse length-scale 
of this parameter. The second kernel we explored was the quasi-periodic kernel introduced by 
\cite{celerite}, which is of the form }
\begin{equation*}
k_{i,j}(\tau) = \frac{B}{2+C}e^{-\tau/L}\left[\cos \left(\frac{2\pi \tau}{P_\textnormal{rot}}\right) + (1+C)\right]
\end{equation*}
{and where $\tau = t_{i} - t_{j}$ is the time-lag. Here $B$ and $C$ are terms that normalize and amplify the 
kernel, whereas $L$ is an exponential decay time-scale and $P_\textnormal{rot}$ is the period of the quasi-periodic 
GP. Finally, we also explored the widely used exp-sine-squared kernel of the form}
\begin{equation*}
k_{i,j}(\tau) = \sigma_{\textnormal{GP}}^2 \exp \left(-\alpha_{\textnormal{GP}} \tau^2 - \Gamma \sin^{2}\left[\frac{\pi \tau}{P_{\textnormal{rot}}}\right]\right),
\end{equation*}
{where $\tau = t_{i} - t_{j}$ is again the time-lag, $\sigma_{\textnormal{GP}}$ is the amplitude of the GP 
component, $\alpha_{\textnormal{GP}}$ is an inverse time-scale for the GP, $\Gamma$ is the amplitude of the 
periodic component of the GP and $P_\textnormal{rot}$ is, again, the period of the quasi-periodic component. 
The priors used for the hyper-parameters of those GP models are listed in Table \ref{tab:priorsGP}.} 

{The first four items in Table \ref{tab:evidenceGP} show the results of this first group of fits performed on 
our data. As can be seen, among those models the best one given the data appears to be the one which 
includes one planet (the transiting one) plus the kernel introduced by \cite{celerite}. This result is 
interesting because the periodic component of the GP is clearly trying to fit for a 4.8-day periodic component 
plus some extra signal in the data in this case, which led us to believe that the best model could be one which has 
two Keplerians (one for the transiting planet and one for the 4.8-day signal) and an additional GP component 
on top of them (i.e., point (3) above). Motivated by this possibility, we performed a second group of fits with 
two Keplerians plus a GP, where we tried the same kernels as for the first group of fits (i.e., with the priors on 
the GP hyperparameters given in Table \ref{tab:priorsGP}). The results of our fits for this second group of fits 
are also presented in Table \ref{tab:evidenceGP} (three last items in the list).}

\begin{table*}
    \centering
    \caption{Priors used for our {fits including GPs.} These were used in conjunction with the priors 
    listed in Table \ref{tab:priors}.}  
    \label{tab:priorsGP}
    \begin{tabular}{lccl} 
        \hline
        \hline
        Parameter name & Prior & Units & Description \\
                \hline
                \hline
        Parameters for the squared exponential kernel & & \\
        ~~~$\sigma_{\textnormal{GP}}$ &$\mathcal{J}(10^{-5},1000)$ & m/s & Amplitude of GP component. \\
        ~~~$\alpha_{\textnormal{GP}}$ &$\mathcal{J}(10^{-5},1000)$ & --- & Inverse length-scale of the GP component. \\
        \hline
        Parameters for the quasi-periodic kernels & & \\
         ~~~$P_{\textnormal{rot}}$ &$\mathcal{N}(4.75,1.00^2)$ & days & Period of the quasi-periodic component. \\ 
        Parameters for \citep{celerite} quasi-periodic kernel & & \\
         ~~~$B$ & $\mathcal{J}(10^{-5},1000)$ & m$^2$/s$^2$ & Amplitude of GP component. \\ 
         ~~~$C$ & $\mathcal{J}(10^{-5},1000)$ & --- & Factor of GP component. \\ 
         ~~~$L$ & $\mathcal{J}(10^{-5},1000)$ & 1/day & Lengthscale of the quasi-periodic component. \\ 
        Parameters for the exp-sine-squared GP & & \\
         ~~~$\sigma_{\textnormal{GP}}$ & $\mathcal{J}(10^{-5},1000)$ & m/s & Amplitude of GP component. \\ 
         ~~~$\alpha_{\textnormal{GP}}$ & $\mathcal{J}(10^{-5},1000)$ & --- & Inverse length-scale of the GP component. \\ 
         ~~~$\Gamma$ & $\mathcal{J}(10^{-5},1000)$ & --- & Amplitude of the sine-squared term in the GP. \\ 
        \hline
    \end{tabular}
\end{table*}

\begin{table}
    \centering
    \caption{Resulting log-evidences (and differences with respect to the model selected as the ``best" model --- in bold) from different model fits to the full photometric and RV datasets with the priors defined in Tables \ref{tab:priors} and \ref{tab:priorsGP} (see text). SE stands for results using a squared-exponential kernel, whereas FM stands for results using the \protect\cite{celerite} quasi-periodic kernel. The value presented for the 
    1 planet + SE GP below corresponds to the model using 
    time as a variable, which was the model that gave the 
    best fit among that class of models. The 2 planet + SE GP model 
    was selected as the best model as is indistinguishable between the other 2-planet GP fits ($|\Delta \ln Z < 1|$), and is the 
    simpler (i.e., has lower number of free parameters) of them.}  
    \label{tab:evidenceGP}
    \begin{tabular}{lcc} 
        \hline
        \hline
        Model & $\ln Z$ & $\Delta \ln Z$ \\
                \hline
                \hline
        2 planets ~~~ & 111,484.42 & -49.0 \\
        1 planet + SE GP ~~~ & 111,505.87 & -27.5 \\
        1 planet + exp-sine-squared QP GP ~~~ & 111,516.60 & -16.8 \\
        1 planet + FM QP GP ~~~ & 111,526.35 & -7.07 \\
        2 planets + exp-sine-squared QP GP ~~~ & 111,533.07 & -0.35 \\
        {2 planets + SE GP} & {111,533.42} & {0}\\
        2 planets + FM QP GP ~~~ & 111,533.61 & 0.19 \\
        \hline
    \end{tabular}
\end{table}

\begin{figure}
   \includegraphics[height=0.75\columnwidth]{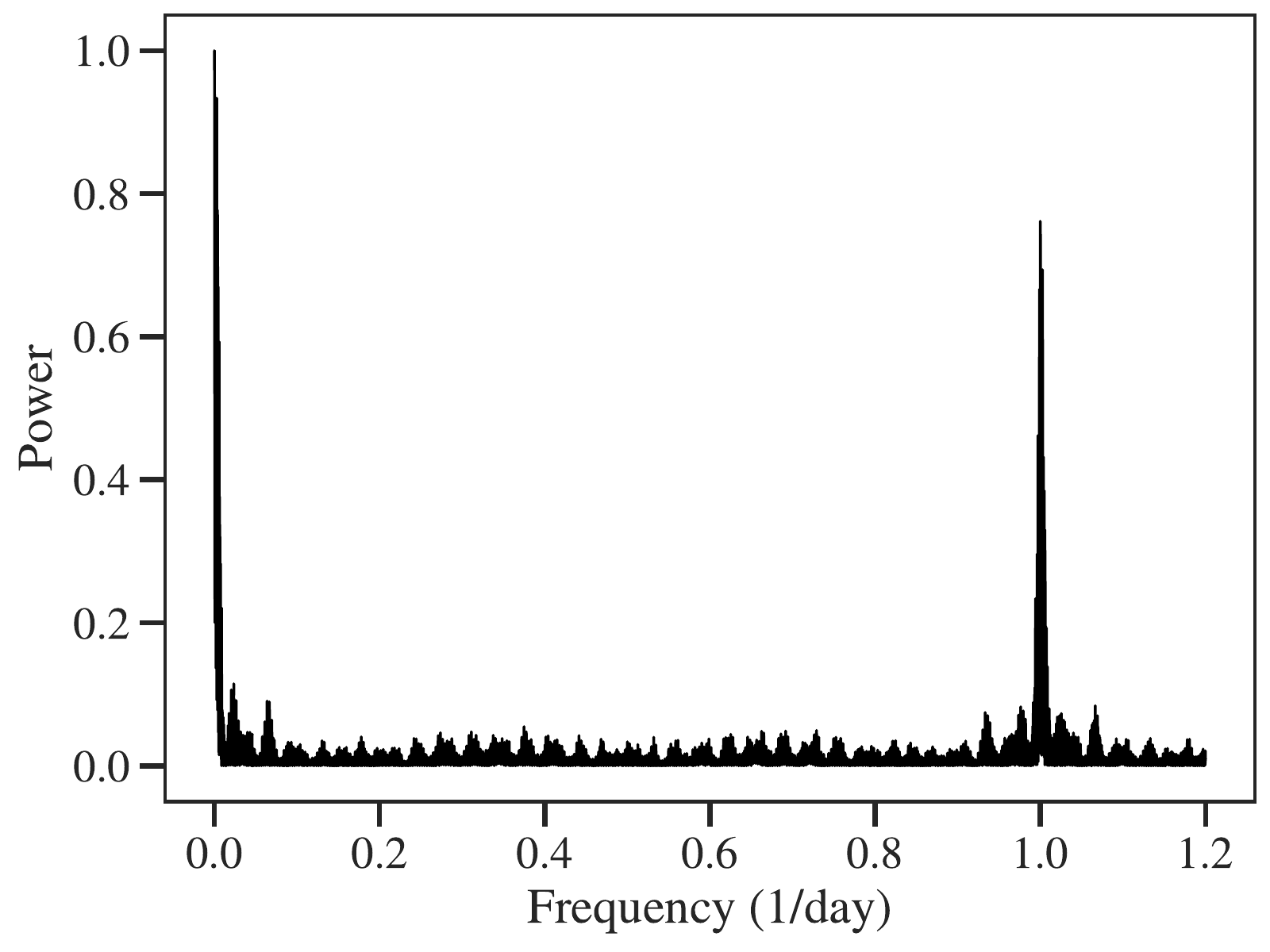}
    \caption{{Window function of our radial-velocity samples. Two peaks emerge in the window shown here, one at 0.001 day$^{-1}$ 
    and another one at 1.00185 day$^{-1}$.}}
    \label{fig:wf}
\end{figure}
\begin{figure*}
   \includegraphics[height=2.0\columnwidth]{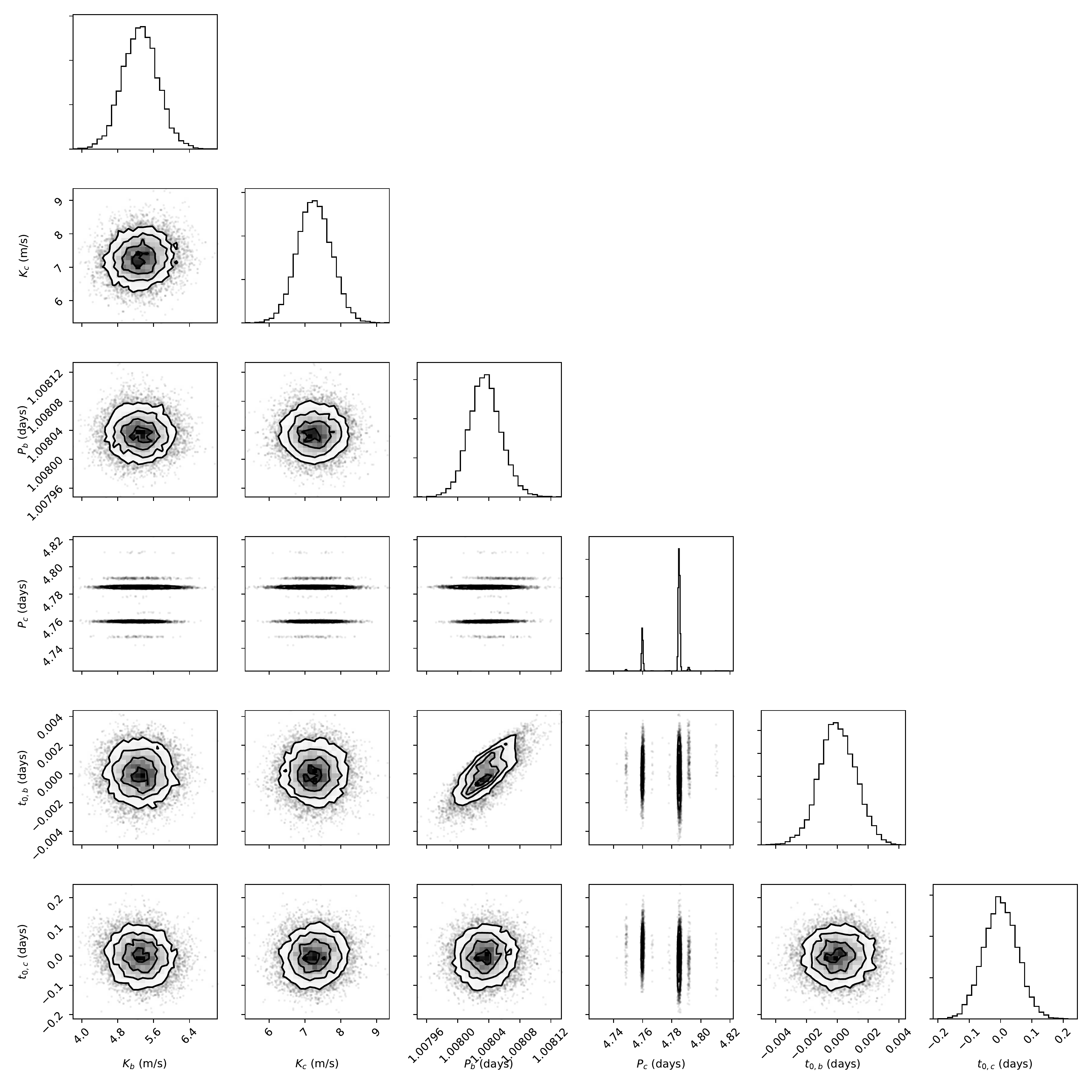}
    \caption{{Corner plot of the posterior distribution of the main parameters of planet b and c, where the multi-modality of the period of planet $c$ is evident. The two main peaks of this multi-modal distribution are located at $4.75983^{+0.00046}_{-0.00043}$ days and 
    at $4.78503^{+0.00056}_{-0.00051}$ days. We note that for the time-of-transit centers, the plotted values are the 
    median-substracted values of the posteriors. This has been substracted for clarity in the corner plot.}}
    \label{fig:corner}
\end{figure*}

As can be seen in Table \ref{tab:evidenceGP}, 
the models with the highest evidences are models 
with two planets and an additional GP component\footnote{{We also tried 3-planet fits, but these show much smaller 
log-evidences than the models presented here. A 3-planet fit with the third component having a log-prior on the period from 5 
to 100 gives a log-evidence worse than the best model presented in this work.}}. At 
face value, the model with the highest evidence is 
the one using the \cite{celerite} quasi-periodic kernel, 
but this model is in practice indistinguishable 
($|\Delta \ln Z < 1|$) from both a fit using an exp-sine-squared kernel and a squared-exponential kernel. Interestingly, the 
quasi-periodic kernels in these 2-planet fits actually provide 
no constraint on the extra residual periodic component --- 
the posterior on the $P_\textnormal{rot}$ parameters only 
rules out periods smaller than about 5-days, and is uniform in 
the rest of the parameter space, which hints that the prescence of any additional 
periodic signal (e.g., activity and/or extra planetary companions) is unlikely 
given our data. In fact, all three fits 
converge to the same posterior parameters for all the orbital 
and physical parameters of the planets in the system. Being 
the 2-planet plus squared-exponential GP the simpler 
of the mentioned fits, we choose this as our best model 
in this case; this model is in turn superior to both the 
1-planet models assuming an extra squared-exponential kernel or a 
quasi-periodic kernel and to the 2-planet fit without a GP 
component. Together with our discussion in 
the previous section that stellar activity indicators show 
no evident peaks in the periodogram at the periods of 
interest, and that the rotation period of the star is 
much longer than the period of interest, we take this 
as evidence that the observed signal is indeed caused 
by a non-transiting planet, to which we refer from now on to 
as \planetnamec. It is interesting to note that our posterior 
distribution for the period of planet \planetnamec\ is 
actually multi-modal with the two main periods being 
at $4.75983^{+0.00046}_{-0.00043}$ days and at $4.78503^{+0.00056}_{-0.00051}$ days. It is under this 
latter period that most of the posterior density is located, {in fact. However, an additional 
piece of evidence that this latter one is the true period of \planetnamec\ comes from examining our 
window function (Figure \ref{fig:wf}). The function shows the expected peak around the solar day (1.0018 day$^{-1}$ in 
our window function), which in turn is propagated also to lower and higher frequencies. In particular, the largest peak 
in our window function in frequency space is at $f_\textnormal{s} = 0.001$ day$^{-1}$. Given a real frequency present 
in the data, thus, aliases of this frequency will emerge at $f_\textnormal{alias} = f_\textnormal{true} \pm m f_\textnormal{s}$, where 
$m$ is an integer, $f_\textnormal{true}$ is the true, underlying frequency embedded in the dataset, $f_\textnormal{alias}$ is 
the generated alias and $f_\textnormal{s}$ is a peak from the window function. Indeed, if the $4.78503$ period is the real period, 
then with $f_\textnormal{s} = 0.001$ day$^{-1}$ this signal should generate aliasing signals at periods of $4.76$ and $4.81$ days, 
both of which we do see in our posterior distribution for the period. If the real period were $4.75983$ days, on the other hand, 
this should give rise to aliasing signals at periods of 4.78 and 4.73 days --- the latter not being present in our posterior 
distribution. To make a quantitative assessment of this, we used the \texttt{AliasFinder} package (Stock \& Kemmer, in prep.)\footnote{\url{https://github.com/JonasKemmer/AliasFinder}}, which implements the procedure for alias finding 
detailed in \cite{DF:2010}. Using this tool with both of these periods yields the same suggestion: that the  $4.78503$ period is the real period, with the $4.75983$ period being an alias. Because of this, the period of $4.78503^{+0.00056}_{-0.00051}$ days} 
is the one we report as our final estimate for the period of this exoplanet. We note that there is no 
strong correlation between the period and any other parameter in Figure \ref{fig:corner} --- this multi-modal nature, however, 
appears {not only on our GP fits but also on our 2-planet white-noise fits, which suggests that, indeed, these appear 
because of the sampling of the data}. Importantly, this bi-modal nature of this period, does not enlarge the uncertainties in the 
other retrieved parameters.

\begin{figure}
   \includegraphics[height=0.8\columnwidth]{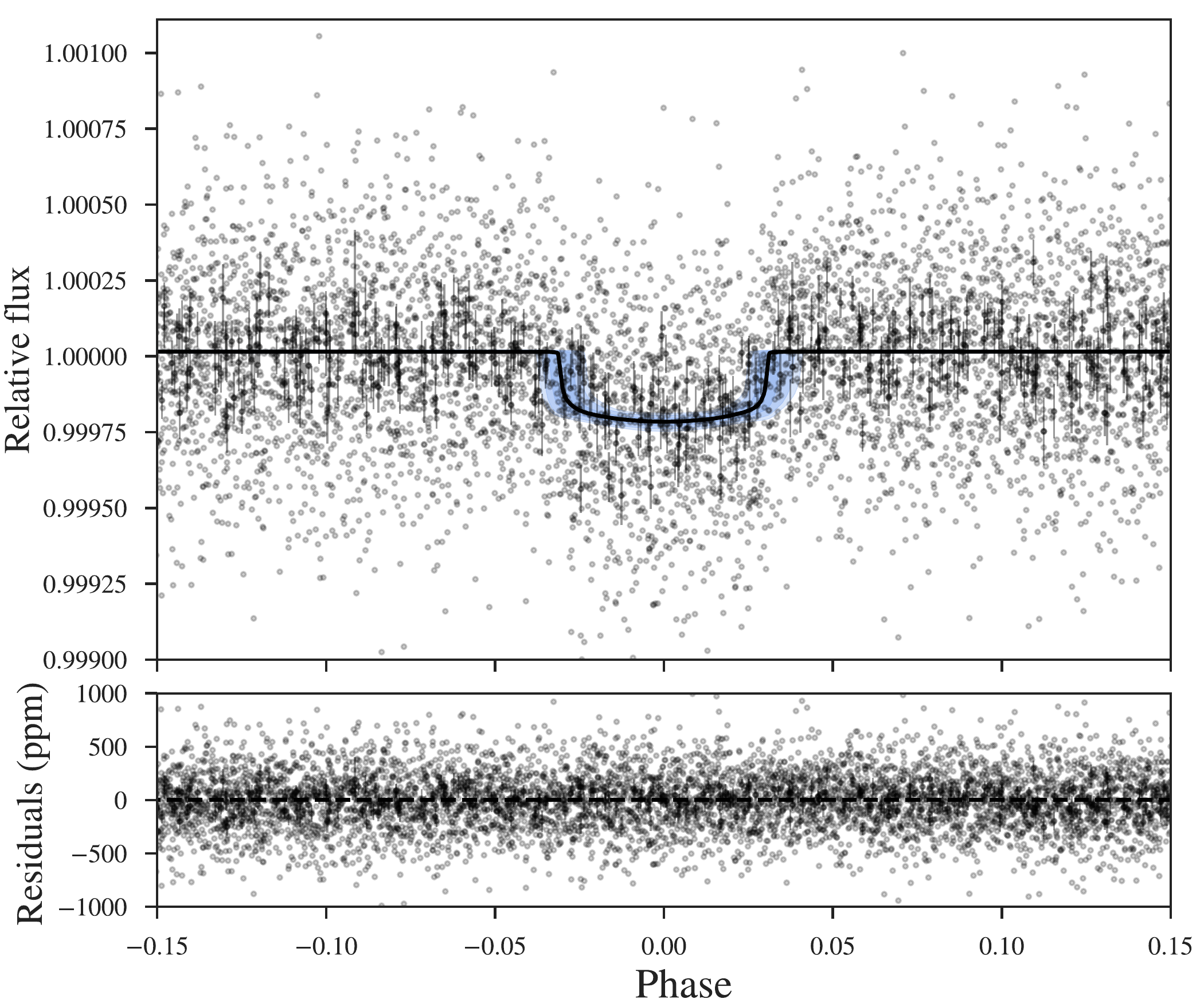}
    \caption{\textit{Top}. TESS photometry phased around the period of \planetnameb\ (grey points; black points with errorbars correspond to 10-point binned photometry shown for illustration). The black line shows the median posterior model given the data, and blue bands denote its 68\%, 95\% and 99\% posterior credibility bands. \textit{Bottom}. Residuals obtained by substracting the data with our median posterior model.}
    \label{fig:photometry}
\end{figure}

\begin{figure*}
   \includegraphics[height=2.05\columnwidth]{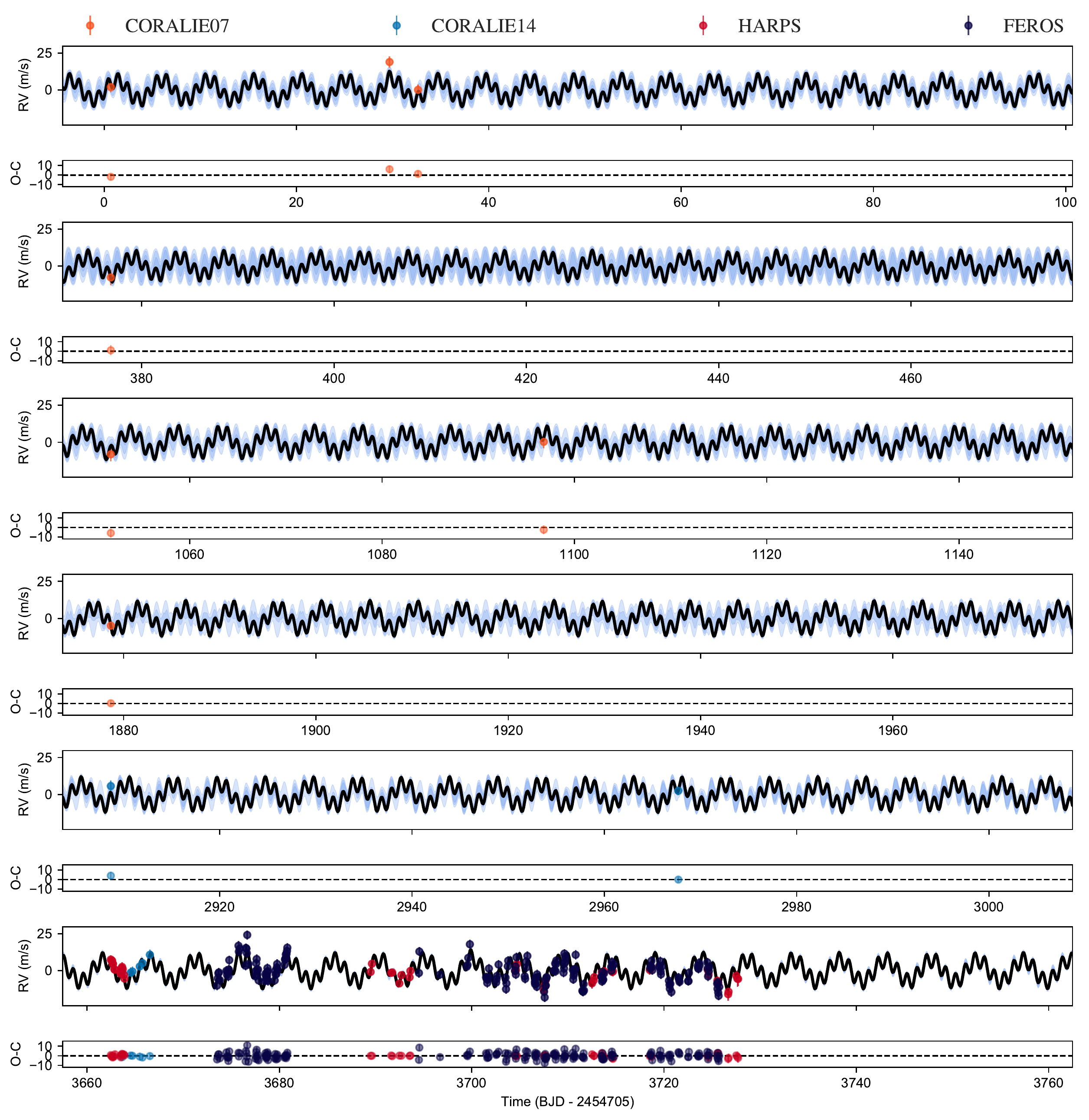}
   \includegraphics[height=0.62\columnwidth]{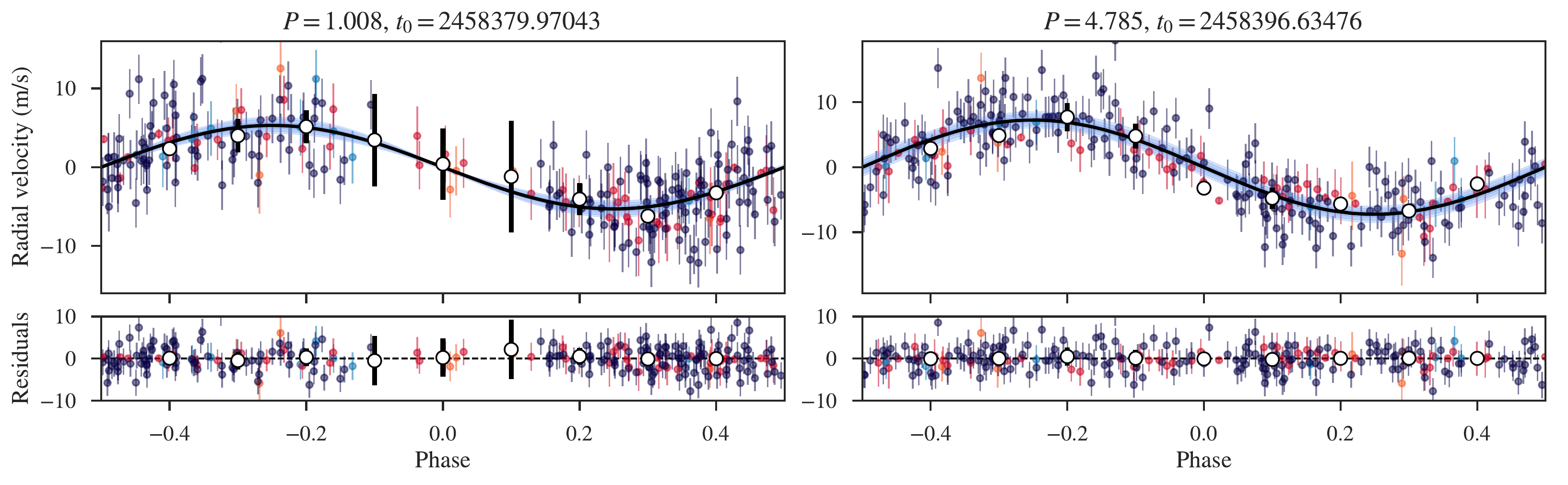}
    \caption{Systemic velocity-substracted radial velocities for the \starname\ system observed by our FEROS (dark blue), HARPS (red) and CORALIE (orange and light blue) observations. The top panel shows the radial velocities as a function of time along with the residuals (O-C) obtained from substracting those with our median posterior model given the data (black lines; blue bands around it denoting 68\%, 95\% and 99\% posterior credibility bands). Note the effects of the sampling of the inner, 1-day period planet, which made us sample almost identical phases on consecutive days. The bottom panel shows the phased radial velocities of \planetnameb\ (bottom left panel) and \planetnamec\ (bottom right panel) with the GP component removed, along with the phased residuals --- white points 
    show binned datapoints in phase for visualization. The same coloring as for the top panels is used for the bottom panels.}
    \label{fig:rvs}
\end{figure*}

\begin{figure}
   \includegraphics[height=0.55\columnwidth]{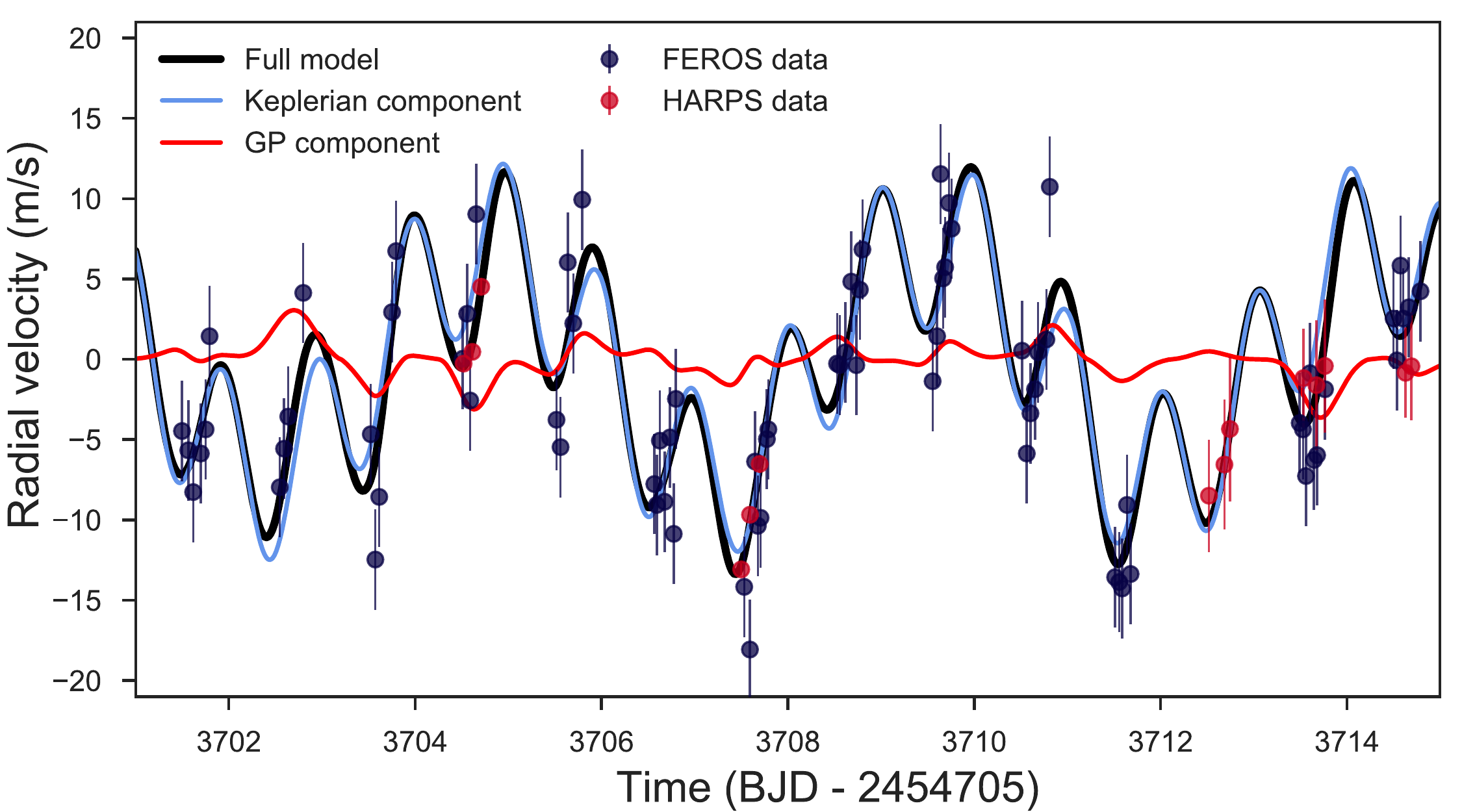}
    \caption{Close-up to the radial-velocity dataset presented in Figure \ref{fig:rvs}, where we show each component of our best-fit model (black line): the keplerian (blue line) and the GP (red line) component.}
    \label{fig:rv-closeup}
\end{figure}

The posterior distribution of the parameters of our best-fit 
model are presented in Table \ref{tab:posteriors} for all the 
parameters except for the eccentricities and the 
jitter terms mentioned above --- for those parameters we 
present upper limits based on the fits performed allowing 
those to vary freely in our \codename\ runs; the corresponding 
posterior modelling of the data is presented in 
Figure \ref{fig:photometry} for the photometry 
and Figure \ref{fig:rvs} for the radial velocities. A close-up 
to the radial velocities showing how each component of our model 
adds to the full signal is presented in Figure \ref{fig:rv-closeup}.

\begin{table}
    \centering
    \caption{Posterior parameters obtained from our \codename\ analysis for \planetnameb\ and \planetnamec}
    \label{tab:posteriors}
    \begin{tabular}{lc} 
        \hline
        \hline
        Parameter name & Posterior estimate$^a$ \\
                \hline
                \hline
        Posterior parameters for \planetnameb & \\[0.1cm]
        ~~~$P_b$ (days) & $1.008035^{+0.000021}_{-0.000020}$ \\[0.1 cm]
        ~~~$t_{0,b}$ (BJD UTC) & $2458379.97043^{+0.0012}_{-0.0012}$ \\[0.1 cm]
        ~~~$\rho_*$ (kg/m$^3$) & $1127.4^{+31.8}_{-31.3}$ \\[0.1 cm]
        ~~~$r_{1,b}$ & $0.783^{+0.022}_{-0.027}$ \\[0.1 cm]
        ~~~$r_{2,b}$ & $0.01453^{+0.00041}_{-0.00042}$ \\[0.1 cm]
        ~~~$K_{b}$ (m/s) & $5.30^{+0.39}_{-0.39}$ \\[0.1 cm]
        ~~~$e_b$ & 0 (fixed$^b$, $<0.24$) \\[0.1 cm]
        \hline
        Posterior parameters for \planetnamec & \\[0.1cm]
        ~~~$P_c$ (days) & $4.78503^{+0.00056}_{-0.00051}$ \\[0.1 cm]
        ~~~$t_{0,c}$ (BJD UTC) & $2458396.635^{+0.054}_{-0.054}$ \\[0.1 cm]
        ~~~$K_{c}$ (m/s) & $7.26^{+0.48}_{-0.47}$ \\[0.1 cm]
        ~~~$e_c$ & 0 (fixed$^b$, $<0.16$)\\[0.1 cm]
        \hline
        Posterior parameters for TESS photometry & \\[0.1cm]
        ~~~$M_{\textnormal{TESS}}$ (ppm) &$-21^{+2.2}_{-2.2}$ \\[0.1 cm]
        ~~~$\sigma_{w,\textnormal{TESS}}$ (ppm) & $111.1^{+4.0}_{-4.0}$ \\[0.1 cm]
        ~~~$q_{1,\textnormal{TESS}}$ & $0.23^{+0.31}_{-0.17}$ \\[0.1 cm]
        ~~~$q_{2,\textnormal{TESS}}$ & $0.43^{+0.35}_{-0.29}$ \\[0.1 cm]
        \hline
        Posterior RV parameters & \\[0.1cm]
        ~~~$\mu_{\textnormal{FEROS}}$ (m/s) & $36131.07^{+0.40}_{-0.40}$ \\[0.1 cm]
        ~~~$\sigma_{w,\textnormal{FEROS}}$ (m/s) & $0.88^{+0.82}_{-0.81}$ \\[0.1 cm]
        ~~~$\mu_{\textnormal{HARPS}}$ (m/s) & $36159.65^{+0.53}_{-0.53}$ \\[0.1 cm]
        ~~~$\sigma_{w,\textnormal{HARPS}}$ (m/s) & 0 (fixed$^b$, $<2.16$) \\[0.1cm]
        ~~~$\mu_{\textnormal{CORALIE07}}$ (m/s) & $36088.8^{+1.8}_{-1.8}$ \\[0.1 cm]
        ~~~$\sigma_{w,\textnormal{CORALIE07}}$ (m/s) & 0 (fixed$^b$, $<8.68$) \\[0.1cm]
        ~~~$\mu_{\textnormal{CORALIE14}}$ (m/s) & $36133.1^{+1.4}_{-1.5}$ \\[0.1 cm]
        ~~~$\sigma_{w,\textnormal{CORALIE14}}$ (m/s) & 0 (fixed$^b$, $<2.51$) \\[0.1cm]
        ~~~$\sigma_{\textnormal{GP}}$ (m/s) & $2.08^{+0.32}_{-0.29}$ \\[0.1 cm]
        ~~~$\alpha_{\textnormal{GP}}$ (1/day$^2$) & $27.1^{+27.0}_{-17.2}$ \\[0.1 cm]
        \hline
        Derived transit parameters for \planetnameb & \\[0.1cm]
        ~~~$R_p/R_*$ & $0.01453^{+0.00041}_{-0.00042}$ \\[0.1 cm]
        ~~~$b = (a/R_*)\cos(i_p)$ & $0.675^{+0.033}_{-0.041}$ \\[0.1 cm]
        ~~~$a_{b}/R_*$ & $3.927^{+0.037}_{-0.037}$ \\[0.1 cm]
        ~~~$i_p$ (deg) & $80.09^{+0.62}_{-0.50}$ \\[0.1 cm]
        ~~~$u_1$  & $0.34^{+0.34}_{-0.23}$ \\[0.1 cm]
        ~~~$u_2$  & $0.063^{+0.35}_{-0.27}$ \\[0.1 cm]
        Derived physical parameters for \planetnameb & \\[0.1cm]
        ~~~$M_{p}$ ($M_\oplus$) & \planetmassb \\[0.1 cm]
        ~~~$R_{p}$ ($R_\oplus$) & \planetradiusb \\[0.1 cm]
        ~~~$\rho_{p}$ (g cm$^{-3}$) & \planetrhob \\[0.1 cm]
        ~~~$g_{p}$ (m s$^{-2}$) & \planetgravb \\[0.1 cm]
        ~~~$a$ (AU) & \planetab \\[0.1 cm]
        ~~~$T_\textnormal{eq}$ (K)$^c$ & \planetTeqb \\[0.1 cm]
        Derived physical parameters for \planetnamec & \\[0.1cm]
        ~~~$M_{p}\sin (i_p)$ ($M_\oplus$) & \planetmassc \\[0.1 cm]
        ~~~$a$ (AU) & \planetac \\[0.1 cm]
        ~~~$T_\textnormal{eq}$ (K)$^c$ & \planetTeqc \\[0.1 cm]
        \hline
        \hline
    \end{tabular}
    \begin{tablenotes}
      \small
      \item $^a$ Errorbars denote the $68\%$ posterior credibility intervals (CI).
      \item $^b$ Limits denote the 95\% upper CI of fits allowing all orbits to be eccentric.
      \item $^c$ Equilibrium temperatures calculated assuming 0 Bond Albedo.
    \end{tablenotes}
\end{table}

As can be observed in Figure \ref{fig:rv-closeup} and from the 
derived inverse length-scale reported in 
Table \ref{tab:posteriors}, the GP-component tries to explain 
a stochastic variation with a typical 
time-scale ($1/\sqrt{2\alpha_\textnormal{GP}}$) of 
$\sim 3$ hours with an amplitude of about $\sim 2$ m/s. It is 
unlikely this is some kind of stellar oscillation, as the 
amplitude of them in radial velocities of stars similar to 
the Sun like \starname\ are about one order of magnitude 
smaller and occur at scales of minutes and not of hours 
\citep[see, e.g.,][]{os:2003,aste:2004}. One possibility is 
that our GP component is modelling instrumental systematics; 
these could be coming mainly from the FEROS dataset, which is 
the dominant source of RVs in our work, for which stability 
at the precision level attained in this work ($\sim 2$ m/s) 
has not been tested so far at such timescales.

The derived physical parameters presented in 
Table \ref{tab:posteriors} for the transiting 
exoplanet \planetnameb\ present a remarkable similarity 
with the benchmark exoplanet 55 Cancri e \citep[a.k.a. Janssen,][]{Fischer:2008,Winn:2011}. According to 
the latest analysis of this latter transiting 
exoplanet by \cite{Bourrier:2018}, 55 Cancri e has a radius of $R_p=1.88\pm 0.03 R_\oplus$ and mass of $M_p = 8.0 \pm 0.3 M_\oplus$, which implies a density of $\rho_p = 6.7 \pm 0.4$ g cm$^{-3}$. Similarly 
\planetnameb\ has a radius of $R_p =$ \planetradiusb$R_\oplus$ and mass of $M_p =$ \planetmassb$M_\oplus$, 
which in turn implies a (larger, {but still consistent at 2-sigma}) density of \planetrhob g cm$^{-3}$. Both exoplanets, thus, 
have statistically indistinguishable masses 
and radii (\planetnameb\ is only $\Delta R_p = 0.135 \pm 0.06 R_\oplus$ smaller than 55 Cancri e, i.e., 
consistent with 0 within $\sim 2$ standard deviations). {In fact, this also applies to their 
irradiation levels as well:} the zero-albedo equilibrium temperature 
of \planetnameb\ is only slightly higher than 
that of 55 Cancri e (only 
$\Delta T_\textnormal{eq} \approx 200$ K 
hotter than 55 Cancri e). Thus, 
\planetnameb\ can 
be thought of as a very similar exoplanet to 55 Cancri e, making it 
almost an analogue in terms of the planetary properties{, given the current data at hand}. 
We discuss the prospects that \planetnameb\ provides for planetary characterization and comparative exoplanetology of 
transiting super-Earth exoplanets in light of this similarity in the next section.

The derived properties for \planetnamec\ are 
exciting as well. The minimum mass for \planetnamec\ 
of $M_p\sin(i_p) =$ \planetmassc$M_\oplus$ suggests a minimum 
mass on the order of that of Neptune. Given the 
transiting nature of \planetnameb, we thus expect the 
inclination of this exoplanet to be not much larger 
than its inner companion, implying a true mass 
of the same order as the one implied by its minimum 
mass in our analysis. 

\subsection{Searching for transits of \planetnamec}
\begin{figure}
   \includegraphics[height=0.75\columnwidth]{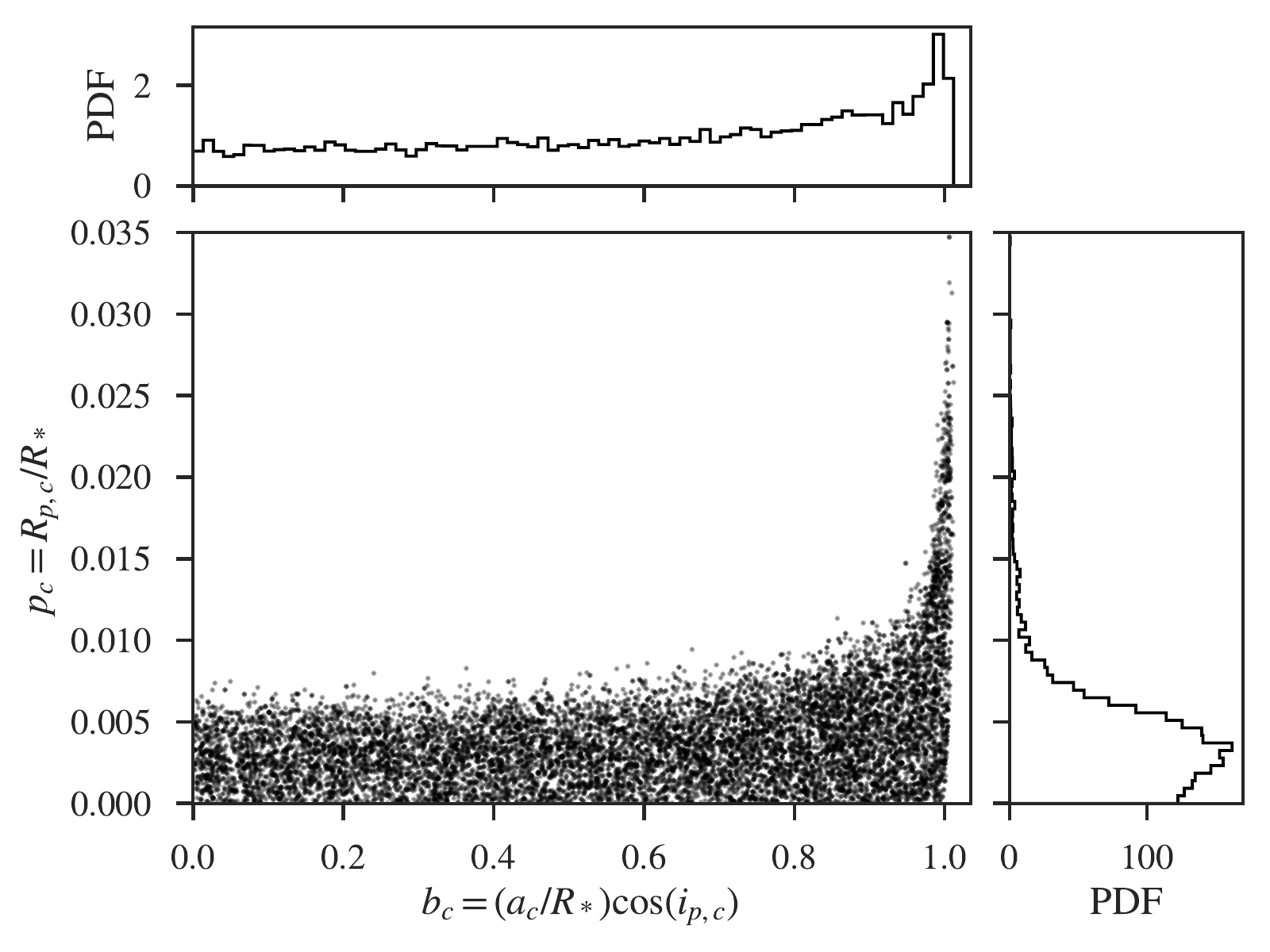}
    \caption{Posterior distribution of the impact parameter ($b_c$) and 
    the planet-to-star radius ratio ($p_c$) for \planetnamec\ given 
    the TESS photometry (central panel). The upper and right-side panels show the marginal distributions of each of those parameters.}
    \label{fig:bp_planetc}
\end{figure}

We used the TESS photometry to search for 
transits of \planetnamec. For this, we performed a \codename\ run with the same priors 
as the ones defined in Table \ref{tab:priors} for \planetnamec. We assumed a circular 
orbit for both exoplanets (as per our result in Section \ref{sec:jointanalysis}) and we added $r_{1,c}$ and 
$r_{2,c}$ as free parameters to \planetnamec\ with the same priors as the corresponding parameters for 
\planetnameb\ to allow a transiting scenario 
for \planetnamec. The resulting \codename\ runs 
with and without a transiting \planetnamec\ with this parametrization significantly favored the non-transiting model 
($\ln Z = 5.4$ in 
favor of this model). Figure \ref{fig:bp_planetc} shows the 
posterior distribution of the impact parameter $b_c = (a_c/R_*)\cos(i_{p,c})$ (where $a_c$ is the semi-major axis of 
planet $c$ and $i_{p,c}$ is the inclination of planet $c$) 
and the planet-to-star radius ratio of the planet, $p_c = R_{p,c}/R_*$, in the case of the joint fit assuming 
\planetnamec\ transits. The marginal 
distribution of the planet-to-star radius ratio implies that 
even if the planet were to transit, about 95\% of the posterior density is bounded to be $p_c<0.014$, i.e., a planetary radius $R_{p,c}<1.7R_\oplus$. At the same time, in this transiting scenario the impact parameter, 
coupled with the tight constraint on the stellar density (and 
hence, on $a_c/R_*$) would imply that 95\% of the posterior density 
of the inclination is above $i_{p_c}>84.82$, implying $\sin(i_{p_c})>0.996$, and hence making the true mass \planetmassc$M_\oplus<M_p<20.05^{+1.36}_{-1.39}M_\oplus$. This would in turn give 
rise to a density for \planetnamec\ about two-times that of \planetnameb, which would imply an extremely dense object. 
The rareness of such an object thus adds to the statistical 
evidence that \planetnamec\ most likely does not transit 
\starname.

\subsection{Secondary eclipses, phase-curve modulations, TTVs}
A search for secondary eclipses and phase-curve modulations of 
either \planetnameb\ or \planetnamec\ turned out to be null in the 
TESS photometry. 
This is not surprising as both reflected and emitted light in 
the TESS bandpass for these exoplanets is expected to be quite low; 
on the order of a couple of ppm for \planetnameb\ and 
a couple tens of ppm for \planetnamec\ depending on its size.

\begin{figure}
   \includegraphics[height=0.7\columnwidth]{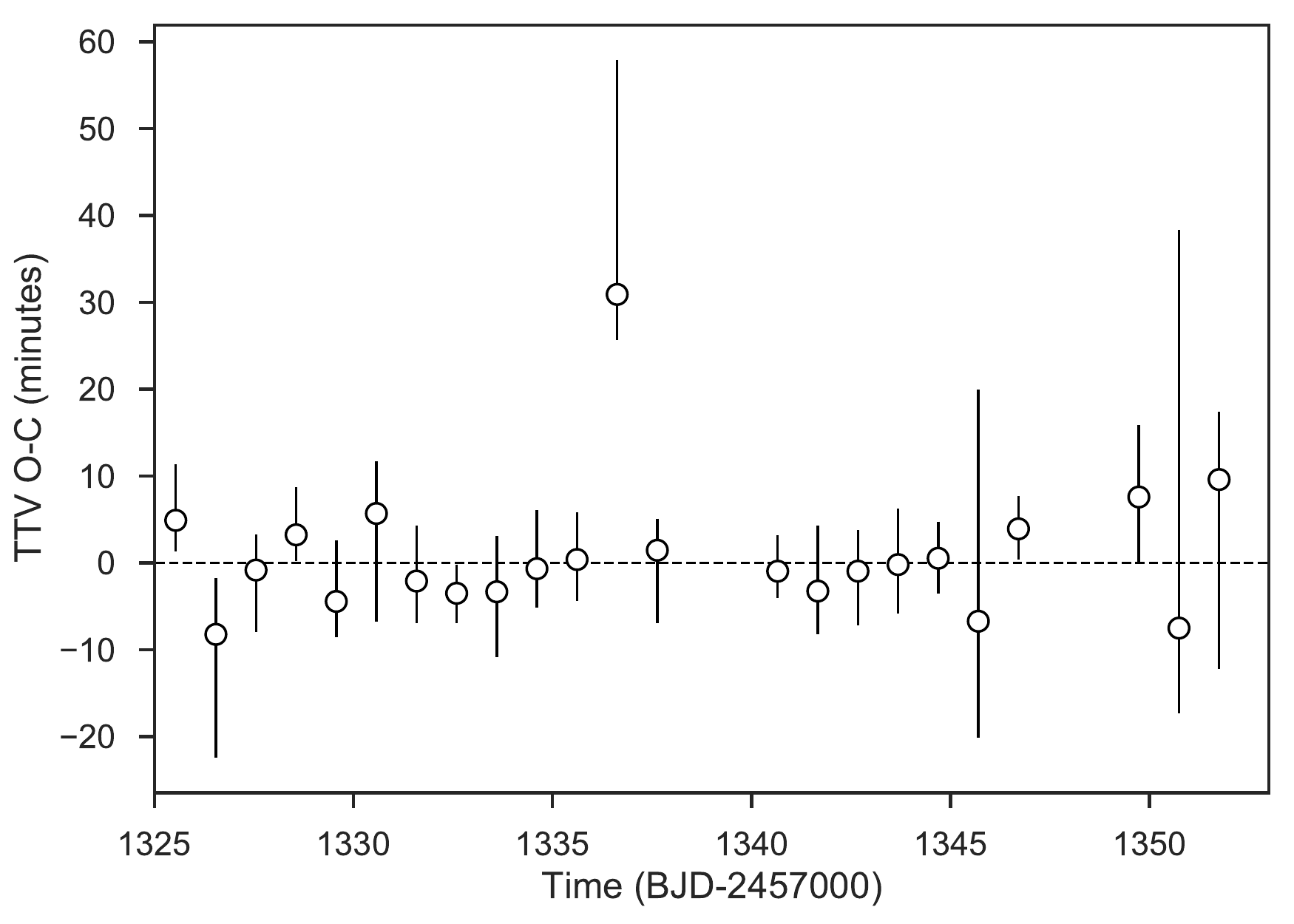}
    \caption{Transit timing variations (TTVs) for \planetnameb\ (i.e., observed minus expected time of transit center as a function of time). No evident variation is observed, putting a limit of $\sim 2$ minutes to any TTV for \planetnameb.}
    \label{fig:ttvs}
\end{figure}

In addition, we performed a search for transit timing variations (TTVs) on the transits of \planetnameb. For this, we used the posterior transit parameters presented in Table \ref{tab:posteriors} as priors for transit fits to the individual 
transits with \codename\ where the time-of-transit center was left as 
a free parameter with a uniform prior between 2 hours before and 2 
hours after of the expected time of transit center. The resulting measured 
TTVs are presented in Figure \ref{fig:ttvs}. As can be seen, there are no evident 
TTVs, except for the 12th transit observed by \textit{TESS}, which appears 
to be half an hour later than expected. However, inspecting this 
portion of the lightcurve there is an evident decrease of flux during 
egress, most likely arising from instrumental effects, which is 
what produces this significant shift in the time of transit. With our 
observations, we can put an upper limit of about 2 minutes over a course of 
27 days to any TTVs impacting the time-of-transit centers of \planetnameb. 
This was expected at least for TTVs generated by \planetnamec\ on 
\planetnameb, for which an order-of-magnitude estimate gives a TTV amplitude on the order of 4 seconds \citep{Holman:2005}.

\section{Discussion}
\label{sec:discussion}
\subsection{The \starname\ system}
\label{discussion:thesystem}

\begin{figure}
   \includegraphics[height=0.8\columnwidth]{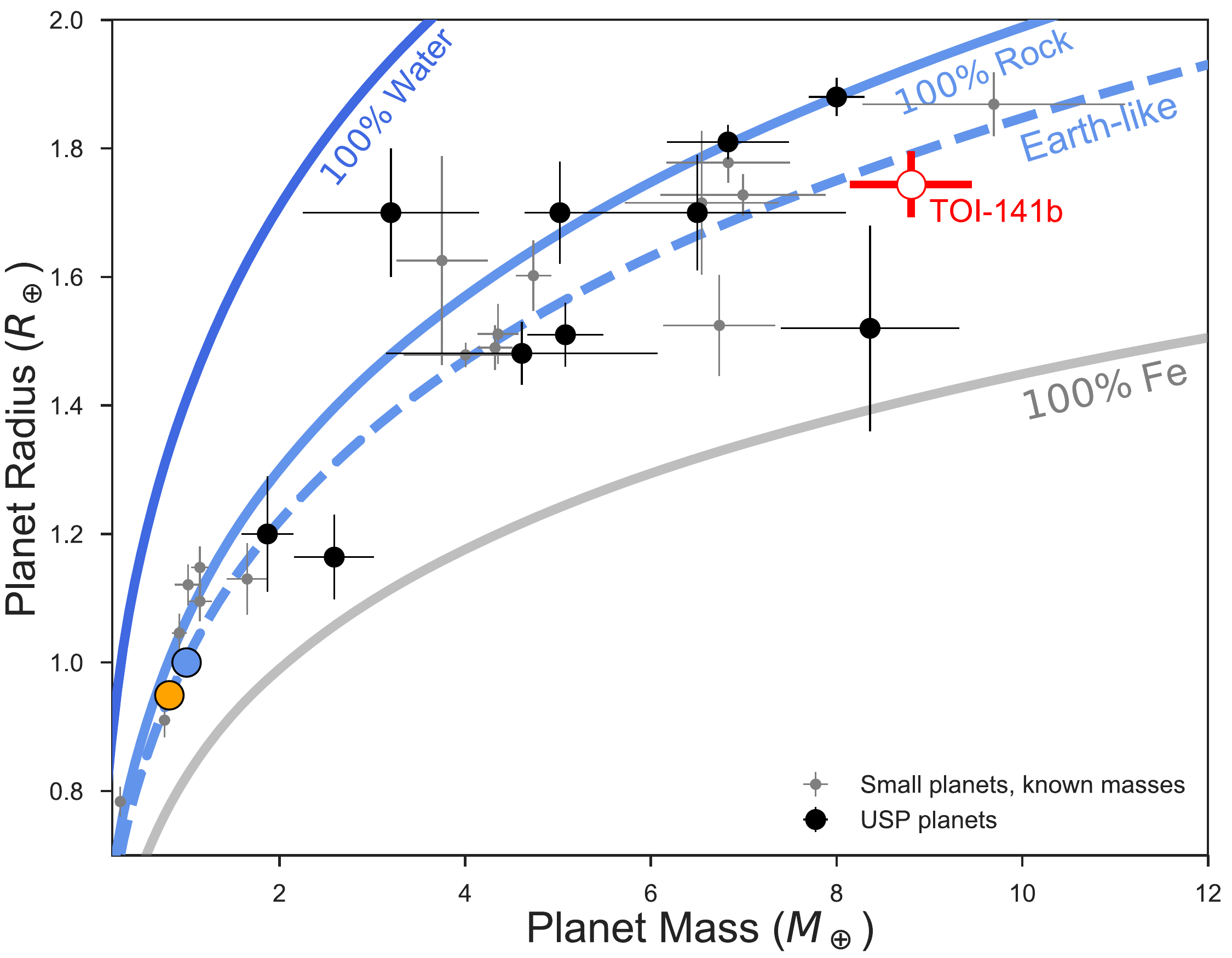}
    \caption{Mass/radius diagram for known exoplanets with sizes smaller than 2 Earth-radii. Black points identify USPs; \planetnameb\ is identified in red. Two-layer models are from \protect\cite{Zeng:2016}; ``Earth-like" here means a composition of 
    30\% Fe and 70\% MgSiO$_3$, whereas ``100 \% 
    Rock" means a composition of 100\% MgSiO$_3$. Earth 
    is identified in this plot as a pale blue circle; 
    the orange circle is Venus.}
    \label{fig:mrd}
\end{figure}
The \starname\ system composed of \planetnameb\ and \planetnamec\ is a very interesting system. On the one hand, \planetnameb, 
as will be shown below in Section \ref{discussion:composition}, 
is a bona-fide ``super-Earth", i.e., a rocky planet 
significantly larger than our home planet. Figure \ref{fig:mrd} compares \planetnameb\ in particular in the mass-radius diagram of exoplanets smaller than 2 Earth-radii (retrieved from \texttt{exoplanets.org}) whose masses and radius are characterized 
at better than 20\%. We plot the two-layer models of 
\cite{Zeng:2016} for illustration.

As can be seen from Figure \ref{fig:mrd}, \planetnameb\ 
appears to have a composition similar to that of 
the Earth according to two-layer models. In fact, among 
super-Earths, it appears this is one of the few 
exoplanets for which we can confidently claim this is the case, making it a very interesting exoplanet. This 
possibility will be discussed in detail in 
Section \ref{discussion:composition}. 

\planetnamec, on the other hand, is most likely a 
short-period Neptune if the mutual inclination 
with \planetnameb\ is not too large. We showed 
that given the data the most plausible scenario for 
\planetnamec\ is that it does not transit the star, 
and thus the maximum inclination of this planet with 
respect of the plane of the sky would be of $i_c < 
\arccos (a/R_*)^{-1}$, or $i_c < 84.829 \pm 0.051$ degrees. This in turn implies that the true mass 
of \planetnamec\ is most likely 
$M_{p,c}>18.54 \pm 0.85 M_\oplus$. We are not able 
to put any constraints on the mutual inclination between 
\planetnameb\ and \planetnamec\ other than this 
upper limit for \planetnamec. 

\subsection{Interior composition of \planetnameb}
\label{discussion:composition}
\subsubsection{Interior characterization: method}

For a detailed interior characterization, we use the probabilistic analysis of \citet{dorn2017generalized} which calculates possible interiors given the observed data (e.g., mass and radius as shown in Figure \ref{fig:mrd}). Besides the data of mass and radius, we are using constraints on the possible bulk composition in terms of refractory elements (e.g., Fe, Mg, Si), which helps to refine interior predictions \citep{dorn2015can}. A proxy for the planet bulk composition is usually taken from the host star's photosphere. Here, measured stellar abundances of \planetnameb\ are (see Section \ref{sec:sab}) [Fe/H] $= -0.04 \pm$ 0.03, [Mg/H] $= -0.04 \pm$ 0.06, and [Si/H] $= -0.03 \pm$ 0.09. Thus, relative stellar abundances of Fe/Si and Mg/Si are similar to the Sun. In brief, our data comprise:
\begin{itemize}
\item Planet masses and radii (Table \ref{tab:posteriors}).
\item Planet effective temperature (Table \ref{tab:posteriors}).
\item Relative stellar abundances of Fe, Si and Mg of the host star.
\end{itemize}

Our assumptions for the interior model are similar to those in \citet{dorn2017generalized}, but we consider a purely rocky planet. We assume an iron core and a silicate mantle, thus \rsolid equals R$_{\rm p}$. The interior parameters are core size $r_{\rm core}$ and mantle composition (i.e., Fe/Si$_{\rm mantle}$, Mg/Si$_{\rm mantle}$).
The prior distributions of the interior parameters are stated in Table \ref{tab:priorinterior}.

\begin{table*}
\caption{Prior ranges for interior parameters.
 \label{tab:priorinterior}}
\begin{center}
\begin{tabular}{lll}
\hline
\hline\noalign{\smallskip}
Parameter & Prior range & Distribution  \\
\noalign{\smallskip}
\hline
\hline\noalign{\smallskip}
Core radius $r_{\rm core}$         & (0.01  -- 1) $r_{\rm core+mantle}$ &uniform in $r_{\rm core}^3$\\
Fe/Si$_{\rm mantle}$        & 0 -- Fe/Si$_{\rm star}$&uniform\\
Mg/Si$_{\rm mantle}$      & Mg/Si$_{\rm star}$ &Gaussian\\
\hline
\end{tabular} 
\end{center}
\end{table*}

Our interior model uses a self-consistent thermodynamic model from \citet[][]{dorn2017generalized}. For any given set of interior parameters, it allows us to calculate the respective mass, radius, and bulk abundances and compare them to the actual observed data. The thermodynamic model comprises the equation of state (EoS) of iron by \citet{bouchet2013ab}, the silicate-mantle model by \citet{connolly2009geodynamic} to compute equilibrium mineralogy and density profiles given the database of  \citet{stixrude2011thermodynamics}. We assume an adiabatic temperature profile within core and mantle.

\subsubsection{Interior characterization: results and discussion}

\begin{figure*}
\centering
 \includegraphics[width = .55\textwidth, trim = 0cm 0cm 0cm 0cm, clip]{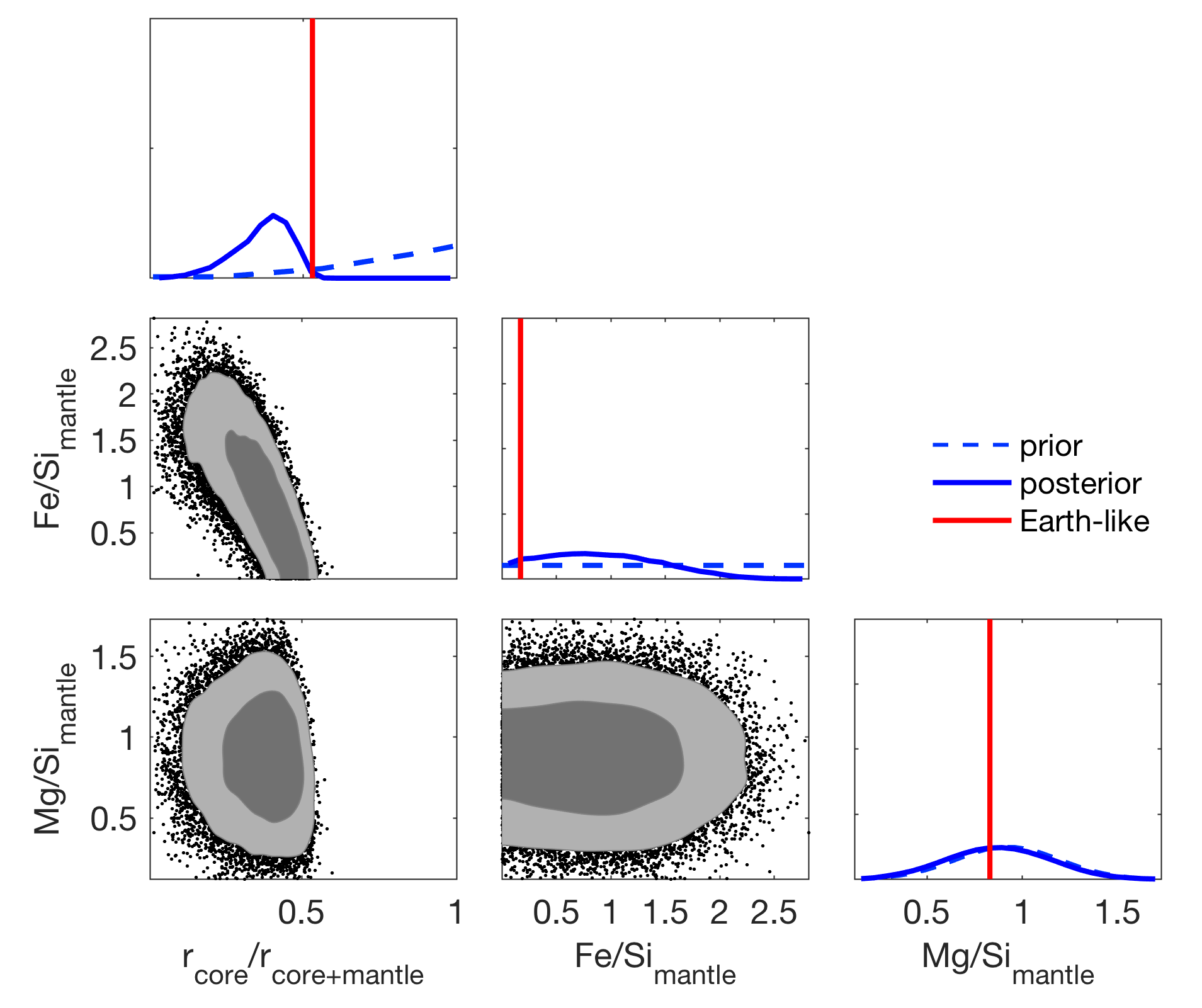}\\
 \caption{Two-and one-dimensional marginalised posteriors of interior parameters: core size ($r_{\rm core}$), and mantle composition (Fe/Si$_{\rm mantle}$ and Mg/Si$_{\rm mantle}$). The prior distribution is shown in dashed, while the posterior distribution is shown in solid lines. An Earth-like interior is shown for reference.}
 \label{plot_interior}
\end{figure*}

Figure \ref{plot_interior} and Table \ref{tableresults} summarize posterior distributions of inferred interior parameters. Given bulk density, the planet is dominated by its rocky interior and might host a very thin terrestrial-type atmosphere only. 
The data of mass, radius, and bulk abundances inform possible core sizes and mantle compositions. Interestingly, the bulk abundance constraints cannot be reconciled with the measured bulk density of $\rho_p=1.66 \rho_\oplus$. This is because the abundance constraint favours Earth-like densities, while \planetnameb's bulk density is higher (see Figure \ref{fig:mrd}). In order to better fit the bulk density, we relaxed the constraint on Fe/Si in a separate scenario and thereby allowed for rocky interiors with larger core mass fractions (Table \ref{tableresults}). Although this scenario can well fit mass and radius, it remains unclear how such iron-rich interiors for massive super-Earths can be formed. The result of a possible iron-rich interior has to be discussed in light of our model assumptions and model uncertainties.  Here, we have assumed pure iron cores for simplicity. The addition of light elements (e.g., O, Si, S, C) in the core can allow for larger cores and thus higher bulk densities, while fitting the measured bulk density. This suggests that the amount of light elements in the core can be constrained by mass, radius, and bulk abundances. Further investigations are underway to understand the importance of light core elements for super-Earths.



Here, we have chosen a rocky interior {\it a priori} and excluded atmospheres to significantly contribute to the planetary radius. We included possible terrestrial-like atmospheres in test runs that showed that possible atmosphere thicknesses are only tiny (0.01 R$_{\rm p}$). Such thin atmospheres cannot be of primordial H/He, since atmospheric escape can efficiently erode thin H/He layer on short time-scales. An atmosphere of H/He is only stable against evaporative loss if it would be significantly thicker than the theoretical minimum threshold-thickness \citep{dorn2018secondary}, which is 0.18 R$_{\rm p}$ for \planetnameb. The threshold-thickness corresponds to the amount of gas (H$_2$) that is lost on short time-scale (here we use 100 Myr).

If this planet has indeed an atmosphere that can be characterized by spectroscopy, this planet would be an interesting target for investigating whether the atmosphere's origin can be informed by the chemical make-up and the extent of the atmosphere. Terrestrial-type atmospheres can  be built during the outgassing of a cooling magma ocean or by volcanism during the long-term evolution of a planet. The rate of volcanism can be very different depending on the convection regime of a planet, e.g., stagnant-lid versus plate tectonics \citep{kite2009geodynamics}. If \planetnameb\ is in stagnant-lid regime, no massive terrestrial-like atmosphere is expected since outgassing rates are very limited for $\gtrapprox$ 8 $M_\oplus$ planets \citep{dorn2018outgassing} despite its partly unconstrained interior structure and composition. A massive atmosphere of volcanic origin could only be present if the planet is in a different convection than stagnant-lid, e.g., plate tectonics. From the variety of modelling studies \citep{valencia2007inevitability,kite2009geodynamics, noack2014plate,korenaga2010likelihood,van2011plate} however, it remains unclear whether Super-Earths can drive plate tectonics or not.

\begin{table*}
\caption{Interior parameter estimates.  One-$\sigma$ uncertainties of the 1-D marginalized posteriors are listed. \label{tableresults}}
\begin{center}
\begin{tabular}{l|lllllll}
\hline\noalign{\smallskip}
interior parameter & all constraints & no Fe/Si constraint & Earth-like value\\ 
\noalign{\smallskip}
\hline\noalign{\smallskip}
$r_{\rm core}$/\rsolid         & $0.38_{-0.11}^{+0.07}$ & $0.41_{-0.12}^{+0.10}$ & 0.53 \\
$\fesima$                      & $0.87_{-0.54}^{+0.60}$ & $1.51_{-0.98}^{+1.46}$ & 0.17 \\
$\mgsima$                      & $0.87_{-0.25}^{+0.26}$ & $0.91_{-0.25}^{+0.25}$ & 0.83 \\
${\rho_p}/\rho_\oplus$         & $1.53_{-0.06}^{+0.07}$ & $1.67_{-0.12}^{+0.13}$ & 1.\\
\hline
\end{tabular} 
\end{center}
\end{table*}

\subsection{Atmospheric characterization of \planetnameb}
Along with Kepler-10b \citep{k10}, Kepler-78b \citep{k78}, K2-141b \citep{k2141} and 55 Cancri e, \planetnameb\ joins the select group of rocky exoplanets that might be optimal targets for 
further atmospheric characterization with current and upcoming facilities. Given \planetnameb\ is so similar to 55 Cancri e, 
an exoplanet that has received particular attention in this front {in recent years} \citep[see, e.g., ][]{Demory:2016,Tsiaras:2016,Angelo:2017,Miguel:2019}, it is important 
to briefly discuss the prospects for atmospheric 
characterization of this newly discovered exoplanet. Among all the 
USPs, \planetnameb\ is the brightest one after 55 Cancri e. However, 
it is 2.4 magnitudes fainter in $K_s$ 
band and $2$ magnitudes fainter in $V$ band than the latter. As such, this 
implies that characterizing the atmosphere of \planetnameb\ will be 
more challenging than the one performed so far for 55 Cancri e with 
known space telescopes such as \textit{Spitzer} and \textit{Hubble}. However, the fact that this provides one of 
the first exoplanets to perform a direct comparison to 
the observational properties of 55 Cancri e, makes this challenge a particularly interesting one to take. For 
the detection of the thermal emission for \planetnameb\ with \textit{Spitzer}, this might involve over 10 transits to 
detect an occultation at 3-sigma confidence. As for transmission, the fainter nature of \planetnameb\ might 
actually help if observations are to be carried out with 
\textit{Hubble}. For 55 Cancri e, spatial scans which left 
larger trails than usual were used in order to compute a 
transmission spectrum with HST/WFC3 \citep[see discussion in][]{Tsiaras:2016}. It is possible 
that this led to precisions 5 to 20\% larger than the 
photon noise, whereas it is known that \textit{Hubble} 
observations can achieve precisions close to 5\% 
the photon-noise for bright stars \citep{Knutson:2014} --- and thus this precision could be achieved for \planetnameb\ with \textit{Hubble}. As such, \planetnameb\ 
might be an excellent target for transmission spectroscopy 
observations with current observatories.

For future \textit{James Webb Space Telescope} (JWST) 
observations, the brightness of \starname\ might 
impact on the type of observations that can be made 
\textit{because the star is too bright}. However, 
observations with different instruments and filters might 
allow to characterize this exoplanet, especially at wavelengths $>2 \mu$m. For example, a wide range of \textit{NIRCam} 
observations are possible to make for \planetnameb\ with a 
range of filters, which implies the thermal emission of this exoplanet might be easily detected with just one JWST 
transit. For transmission, \textit{NIRISS+SOSS} 
observations will be possible for wavelengths 
$\gtrapprox 1.5\mu$m where the instrument saturation 
falls for magnitudes brigther than $J \sim 7$, allowing to 
target a wide range of possible molecular features for 
this exoplanet. In summary, thus, \planetnameb\ could be 
a prime target for JWST transit and occultation observations 
of hot super-Earths.

\section{Conclusions}
In this work, we have presented \planetnameb, a hot 
Super-Earth orbiting in a 1-day period around the G-type star HD 213885 --- the second 
brightest star known to host an ultra-short period exoplanet. The exoplanet was detected by 
\textit{TESS} photometry and later confirmed 
and further characterized using precise RV 
observations with the CORALIE, HARPS and 
FEROS spectrographs. Our observations reveal 
that \planetnameb\ has a rocky bulk composition, converting this exoplanet into a bona-fide 
super-Earth: a rocky planet with a bulk composition similar (although enhanced in iron) to Earth. 
In addition, our precise radial-velocity measurements reveal the presence 
of an additional Neptune-mass exoplanet, \planetnamec, on a 4.78-day orbit which does 
not show transits in the TESS photometry.

\planetnameb\ is an interesting exoplanet from 
the perspective of atmospheric characterization 
of hot super-Earths and especially to be 
compared with 55 Cancri e, for which 
\planetnameb\ is a very similar exoplanet. 
Characterization of this exoplanet with 
both present (e.g., HST, Spitzer) and future 
(e.g., JWST) space-based facilities might 
help unveil the nature of the atmospheres of 
these kind of exoplanets, allowing to 
kickstart comparative exoplanetology of 
hot super-Earths.

\label{sec:conclusions}

\section*{Acknowledgements}
Funding for the TESS mission is provided by NASA's Science Mission directorate. We acknowledge the use of TESS Alert data, which is currently in a beta test phase, from pipelines at the TESS Science Office and at the TESS Science Processing Operations Center. This research has made use of the Exoplanet Follow-up Observation Program website, which is operated by the California Institute of Technology, under contract with the National Aeronautics and Space Administration under the Exoplanet Exploration Program. Resources supporting this work were provided by the NASA High-End Computing (HEC) Program through the NASA Advanced Supercomputing (NAS) Division at Ames Research Center for the production of the SPOC data products. N.E.\ would like to thank the Gruber Foundation for its generous support to this research. 
R.B., A.J., and F.R\ acknowledge support from the Ministry for the Economy, Development, and Tourism's Programa Iniciativa Cient\'{i}fica Milenio through grant IC\,120009, awarded to the Millennium Institute of Astrophysics (MAS).
R.B.\ acknowledges additional support from FONDECYT Postdoctoral Fellowship Project 3180246.
A.J.\ acknowledges additional support from FONDECYT project 1171208.
JSJ acknowledges support by FONDECYT grant 1161218 and partial support from CONICYT project 
Basal AFB-170002. C.A.G. acknowledges support from
CONICYT FONDECYT Postdoctoral Fellowship Project 3180668. DJA acknowledges support from the STFC via an Ernest Rutherford Fellowship (ST/R00384X/1). X.D., H.G., C.L., L. D. N, F. P., S. U., O. T., M. M. and D. S. acknowledge Swiss National Science Foundation (SNSF) for the continuous support of the Swiss EULER-Telescope facility. P.J.W.\ is supported by an STFC consolidated grant (ST/P000495/1). 
SCCB acknowledges support from FEDER - Fundo Europeu de Desenvolvimento Regional funds through the COMPETE 2020 - Operacional Programme for Competitiveness and Internationalisation (POCI), and by Portuguese funds through FCT - Fundacao para a Ciencia e a Tecnologia in the framework of the project POCI-01-0145-FEDER-028953 and the Investigador FCT contract IF/01312/2014/CP1215/CT0004. N.S.C. acknowledges the support by FCT - Fundacao para a Ciencia e a Tecnologia through national funds and by FEDER through COMPETE2020 - Programa Operacional Competitividade e Internacionalizacao by these grants: UID/FIS/04434/2013 \& POCI-01-0145-FEDER-007672; PTDC/FIS-AST/28953/2017 \& POCI-01-0145-FEDER-028953 and PTDC/FIS-AST/32113/2017 \& POCI-01-0145-FEDER-032113. V.A. acknowledges support from FCT through Investigador FCT contract IF/00650/2015/CP1273/CT0001. Work by JNW was supported by Heising-Simons foundation. T.D. acknowledges support from MIT’s Kavli Institute as a Kavli postdoctoral fellow. This work was made possible thanks to 
ESO Projects 0101.C-0510(C) (PI: A. Jord\'an), 1102.C-0249(A) (PI: D. Armstrong), 0102.C-0525(A) (PI: M D\'iaz) and 0102.A-9006(A) (PI: P. Sarkis). C.D. acknowledges the support of the Swiss National Foundation under grant PZ00P2\_174028, and that this work was in part carried out within the frame of the National Center for Competence in Research {\it PlanetS}. K.G.H. is supported by the Polish
National Science Center through grant no. 2016/21/B/ST9/01613

\appendix
\section{Author affiliations}
\label{aa}
$^{1}$ Space Telescope Science Institute, 3700 San Martin Drive, Baltimore, MD 21218, USA.\\
$^{2}$ Max-Planck-Institut f\"ur Astronomie, K\"onigstuhl 17, 69117 Heidelberg, Germany.\\
$^{3}$ Center of Astro-Engineering UC, Pontificia Universidad Cat\'olica de Chile.\\
$^{4}$ Instituto de Astrof\'isica, Facultad de F\'isica, Pontificia Universidad Cat\'olica de Chile.\\
$^{5}$ Facultad de Ingenier\'ia y Ciencias, Universidad Adolfo Ib\'a\~nez, Av.\ Diagonal las Torres 2640, Pe\~nalol\'en, Santiago, Chile.\\
Av. Vicu\~na Mackenna 4860, 782-0436 Macul, Santiago, Chile.\\
$^{6}$ Millennium Institute for Astrophysics, Santiago, Chile.\\
$^{7}$ University of Zurich, Institut of Computational Sciences,
University of Zurich, Winterthurerstrasse 190, CH-8057, Zurich, Switzerland.\\
$^{8}$ Departamento de Astronom\'ia, Universidad de Chile, Camino El Observatorio 1515, Las Condes, Santiago, Chile.\\
$^{9}$ Departamento de Ciencias F\'isicas, Facultad de Ciencias Exactas, Universidad Andr\'es Bello, Las Condes, RM, Santiago, Chile.\\
$^{10}$ NASA Ames Research Center, Moffett Field, CA, 94035.\\
$^{11}$ SETI Institute, Moffett Field, CA  94035, USA.\\
$^{12}$ Center for Astrophysics | Harvard \& Smithsonian, 60 Garden Street, Cambridge, Massachusetts 02138, USA.\\
$^{13}$ Department of Physics, University of Warwick, Gibbet Hill Road, Coventry CV4 7AL.\\
$^{14}$ Centre for Exoplanets and Habitability, University of Warwick, Gibbet Hill Road, Coventry CV4 7AL.\\
$^{15}$  Instituto de Astrof\'isica e Ci\^encias do Espa\c{c}o, Universidade do Porto, CAUP, Rua das Estrelas, PT4150-762 Porto, Portugal.\\
$^{16}$ Depto. Astrof\'{\i}isica, Centro de Astrobiolog\'{\i}a (CSIC-INTA), ESAC campus, Camino Bajo del Castillo s/n, 28692 Villanueva de la Ca\~nada
Madrid, Spain.\\
$^{17}$ Observatoire de l'Universit\'{e} de Gen\`{e}ve, 51 chemin des Maillettes, 1290 Versoix, Switzerland.\\
$^{18}$ European Southern Observatory, Alonso de Cordova 3107, Vitacura Casilla 19001, Santiago 19, Chile.\\
$^{19}$ Departamento de F\'isica e Astronomia, Faculdade de Ci\^encias, Universidade do Porto, Rua do Campo Alegre, 4169-007 Porto, Portugal.\\
$^{20}$ Department of Physics and Kavli Institute for Astrophysics
and Space Science, Massachusetts Institute of Technology, Cambridge, MA 02139, USA.\\
$^{21}$ Department of Astrophysical Sciences, Princeton University,
Princeton, NJ 08544, USA.\\
$^{22}$ Department of Physics and Astronomy, University of Louisville, Louisville, KY 40292, USA.\\
$^{23}$ Computational Engineering and Science Research Centre, University of Southern Queensland, Toowoomba, QLD, 4350, Australia.\\
$^{24}$ Mt. Stuart Observatory, New Zealand.\\
$^{25}$ Dept.\ of Physics \& Astronomy, Swarthmore College, Swarthmore PA 19081, USA.\\
$^{26}$ Nicolaus Copernicus Astronomical Center, Polish Academy of Sciences, ul. Rabia\'{n}ska 8, 87-100 Toru\'{n}, Poland.\\
$^{27}$ Cerro Tololo Inter-American Observatory, National Optical Astronomical Observatory, Casilla 603, La Serena 1700000, Chile.\\
$^{28}$ Dunlap Institute for Astronomy and Astrophysics, University of Toronto, Ontario M5S 3H4, Canada.\\
$^{29}$ Department of Physics and Astronomy, University of North Carolina at Chapel Hill, Chapel Hill, NC 27599-3255, USA.\\
$^{30}$ Noqsi Aerospace Ltd, 15 Blanchard Ave., Billerica, MA 01821, USA.\\
$^{31}$ Exoplanets and Stellar Astrophysics Laboratory, Code 667, NASA Goddard Space Flight Center, Greenbelt, MD 20771, USA.\\
$^{32}$ Millenium Engineering.



\bibliographystyle{mnras}
\bibliography{paperbib}

\bsp	
\label{lastpage}
\end{document}